%% file: main.tex
\documentclass[%
 twocolumn, 
superscriptaddress,
 amsmath,amssymb,
 aps, 
prb,
]{revtex4-2}
\usepackage{amsmath,amssymb}
\usepackage{physics}
\usepackage[dvipdfmx]{graphicx}
\usepackage{subfigmat}
\usepackage{xparse}
\usepackage{xstring}
\usepackage{dcolumn}
\usepackage{bm}
\usepackage{mathtools}
\usepackage{xcolor}
\usepackage{hyperref}
\hypersetup{colorlinks=true,breaklinks,linkcolor=blue,urlcolor=blue,citecolor=blue}
\usepackage[normalem]{ulem}

\begin{document}

\title{Thermal Hall sign reversal from competing magnons and magnon pairs in pyrochlore spin-nematic magnets}

\author{Tetsuya Sato}
\affiliation{Institute for Solid State Physics, University of Tokyo, Kashiwa, 277-8581, Japan}

\author{Emika Takata}
\affiliation{Institute for Solid State Physics, University of Tokyo, Kashiwa, 277-8581, Japan}

\author{Masahiro Sato}
\affiliation{Department of Physics, Chiba University, Chiba 263-8522, Japan}

\date{\today}

\begin{abstract}
Bound magnon pairs are central to spin-nematic order, yet experimentally distinguishing them from single magnons remains challenging. We show that their Berry-curvature-driven heat transport provides a characteristic signature in the thermal Hall response. For a spin-$3/2$ pyrochlore antiferromagnet with ferromagnetic biquadratic exchange in the magnetic-field-induced fully polarized phase, we construct single-magnon bands together with magnon-pair bands, include the Dzyaloshinskii--Moriya (DM) interaction, and evaluate their intrinsic thermal Hall conductivities. The lowest magnon and magnon-pair bands carry Berry curvatures of opposite sign. Because the gaps of the two quasiparticles respond differently to magnetic field, their competing contributions can drive both magnetic-field- and temperature-induced sign reversals of the thermal Hall conductivity. Estimates for chromium spinel oxides indicate that the effect can approach experimentally accessible magnitudes. We also find that the hybridization between the one- and two-magnon sectors induced by the DM interaction preserves this qualitative signature. These results suggest that a broad class of frustrated magnets with both magnons and magnon pairs provides a possible route to tuning the sign of the thermal Hall conductivity and probing bound magnon pairs through thermal transport experiments.
\end{abstract}

\maketitle

\begin{figure}[btp]
	\centering
    \includegraphics[width=0.9\linewidth]{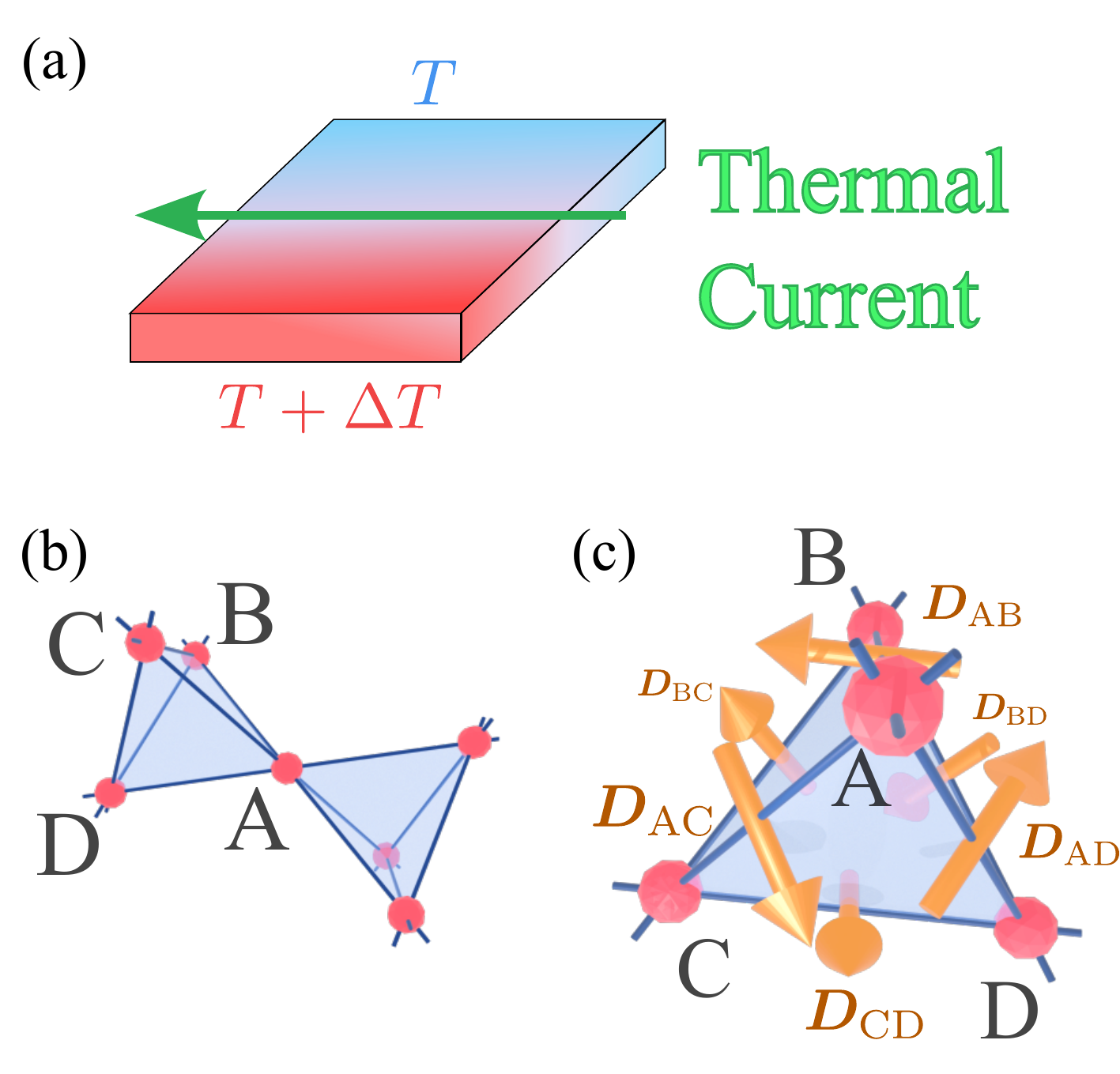}
	\caption{(a) Schematic of the magnon/magnon-pair thermal Hall effect.
    (b) Pyrochlore lattice. The tetrahedra share vertices and form a three dimensional network. Symbols A, B, C and D denote the four sites in a tetrahedron. 
    (c) Four sites contained in the unit cell of the pyrochlore lattice and DM vectors allowed on the nearest-neighboring bonds in the pyrochlore lattice.}\label{fig:pyrochlore}
\end{figure}

\section{Introduction}
As in electronic systems, Berry curvature can give rise to transverse transport even in charge-neutral bosonic bands.
Magnons (spin-wave) are a typical bosonic excitation that appears in a broad class of ordered magnetic insulators. 
In such magnetic insulators, the Dzyaloshinskii–Moriya (DM) interaction is a major source of the complex hopping amplitudes that generate nontrivial band geometry. 
A temperature gradient can therefore drive a transverse magnon heat current without mobile charge carriers as shown in Fig.~\ref{fig:pyrochlore}(a). 
This magnon thermal Hall effect was first predicted theoretically~\cite{katsura2010}, observed in the pyrochlore ferromagnet Lu$_2$V$_2$O$_7$~\cite{onose2010}, and formulated in terms of the Berry curvature and orbital motion of magnon wave packets~\cite{Matsumoto2011,Matsumoto2011_2}. 
Research on topological magnons and their thermal transport has since become a broad field~\cite{LI20211,mcclarty2022}.

Most studies of their thermal transport have focused on one-magnon excitations.
In frustrated magnets, however, attractive interactions between magnons can stabilize two- and multimagnon bound states~\cite{Shannon2006,Momoi2006,Vekua2007,Kecke2007,Hikihara2008,Sudan2009,takata2015nematicorderingpyrochloreantiferromagnets,Takata_Dthesis,mook2023}.
Such bound states are difficult to identify experimentally: A two-magnon bound state (magnon pair) changes $S^z_{\mathrm{tot}}$ by two and is thus inaccessible to probes that couple linearly to a single spin operator.
It is therefore desirable to find transport phenomena that directly reflect the presence of bound magnon pairs~\cite{Hirobe2019}. 
Moreover, the transport of the magnon pairs itself is not well-understood, and the pairs can be a nice candidate for nontrivial spin or heat transport.

Magnon pairs are known to frequently emerge in magnetic-field-induced fully polarized states of frustrated magnets~\cite{Shannon2006,Momoi2006,Vekua2007,Kecke2007,Hikihara2008,Sudan2009,takata2015nematicorderingpyrochloreantiferromagnets,Takata_Dthesis,mook2023}. 
As we mentioned above, an attractive force between magnons often appears in such a forced ferromagnetic phase and it can create the bound magnon pairs. If we decrease the value of applied magnetic field and the excitation gap of magnon pairs vanishes before the magnon gap closing, the Bose-Einstein condensate (BEC) of the magnon pairs occurs. This BEC usually corresponds to the emergence of spin-nematic (spin quadrupolar) order~\cite{Shannon2006,Momoi2006,Zhitomirsky_2010,andreev1984spin,Chubukov1991,penc2004,Tsunetsugu2006,Shannon2010,Ueda2009,Sato2013,Penc2011}. 
Therefore, the formation of magnon pairs is deeply associated with the spin-nematic phase. 

From these arguments, in this paper, we study the thermal Hall conductivity in the magnetic-field-induced fully polarized phase of a spin-$3/2$ pyrochlore antiferromagnet [see Fig.~\ref{fig:pyrochlore}(b)] with ferromagnetic biquadratic exchange interaction~\cite{penc2004,Shannon2010}, where single magnons and bound magnon pairs coexist as the low-energy excitations~\cite{takata2015nematicorderingpyrochloreantiferromagnets}. 
In candidates of pyrochlore antiferromagnets such as ZnCr$_2$O$_4$~\cite{Kemei_2013,lee2000} and HgCr$_2$O$_4$~\cite{ueda2006,kimura2011}, it is known that the Hamiltonian can possess a DM interaction [Fig.~\ref{fig:pyrochlore}(c)] and therefore it is expected to induce complex hoppings and Berry curvature for magnons and magnon pairs~\cite{Elhajal2005,onose2010}.  

We calculate the intrinsic thermal Hall conductivities of the magnon and magnon pairs in the pyrochlore antiferromagnet models with a DM interaction and find that their lowest bands produce contributions of opposite sign. 
A central feature is their different Zeeman shifts: A single magnon changes $S^z_{\mathrm{tot}}$ by one, whereas a magnon pair changes it by two. If we alter the magnetic field, the gap of magnon pairs changes twice as fast as the gap of magnons. 
The magnetic field therefore tunes their relative gaps, so that the total thermal Hall conductivity can reverse sign as a function of temperature or magnetic field.
This characteristic temperature and field dependence provides a possible transport signature of coexisting magnons and bound magnon pairs. 
It should be emphasized that this sign reversal is not accompanied by a dramatic change, such as a phase transition.

The above argument is based on the assumption that total $S^z_{\mathrm{tot}}$ is the conserved quantity, and magnons and magnon pairs are distinguishable as different quasiparticles with different values of $S^z$. 
However, magnetic anisotropies including DM interactions often break the U(1) spin-rotation symmetry, in which magnons and magnon pairs are no longer distinguishable and they are generally hybridized. Even in such symmetry-breaking case, we show that the above scenario of the sign change of thermal Hall conductivity still holds by treating the DM interaction perturbatively. 
We emphasize that the scenario of the sign reversal is not limited to the pyrochlore magnets, and it can hold in a broad class of frustrated magnets with both magnons and magnon pairs (see Sec.~\ref{sec:other_candidates}). 

This paper is organized as follows.
In Sec.~\ref{sec:dispersion}, we derive the magnon and magnon-pair dispersion relations in the pyrochlore magnets and discuss their hybridization by the DM interaction.
In Sec.~\ref{sec:Berry}, we present the Berry curvatures of the magnon and magnon-pair bands.
In Sec.~\ref{sec:thermal_hall}, we briefly review the bosonic thermal Hall formula.
In Sec.~\ref{sec:Numerical}, we quantitatively estimate the thermal Hall conductivities of the magnon and magnon-pair bands and identify the sign reversal and its dependence on temperature and magnetic field.
The experimental feasibility of our prediction is discussed in Sec.~\ref{sec:Experiment_Feasibility}, and candidate systems beyond pyrochlore magnets are discussed in Sec.~\ref{sec:other_candidates}.
Section~\ref{sec:conclusion} concludes the paper.

\begin{figure}
    \centering
    \includegraphics[width=\linewidth]{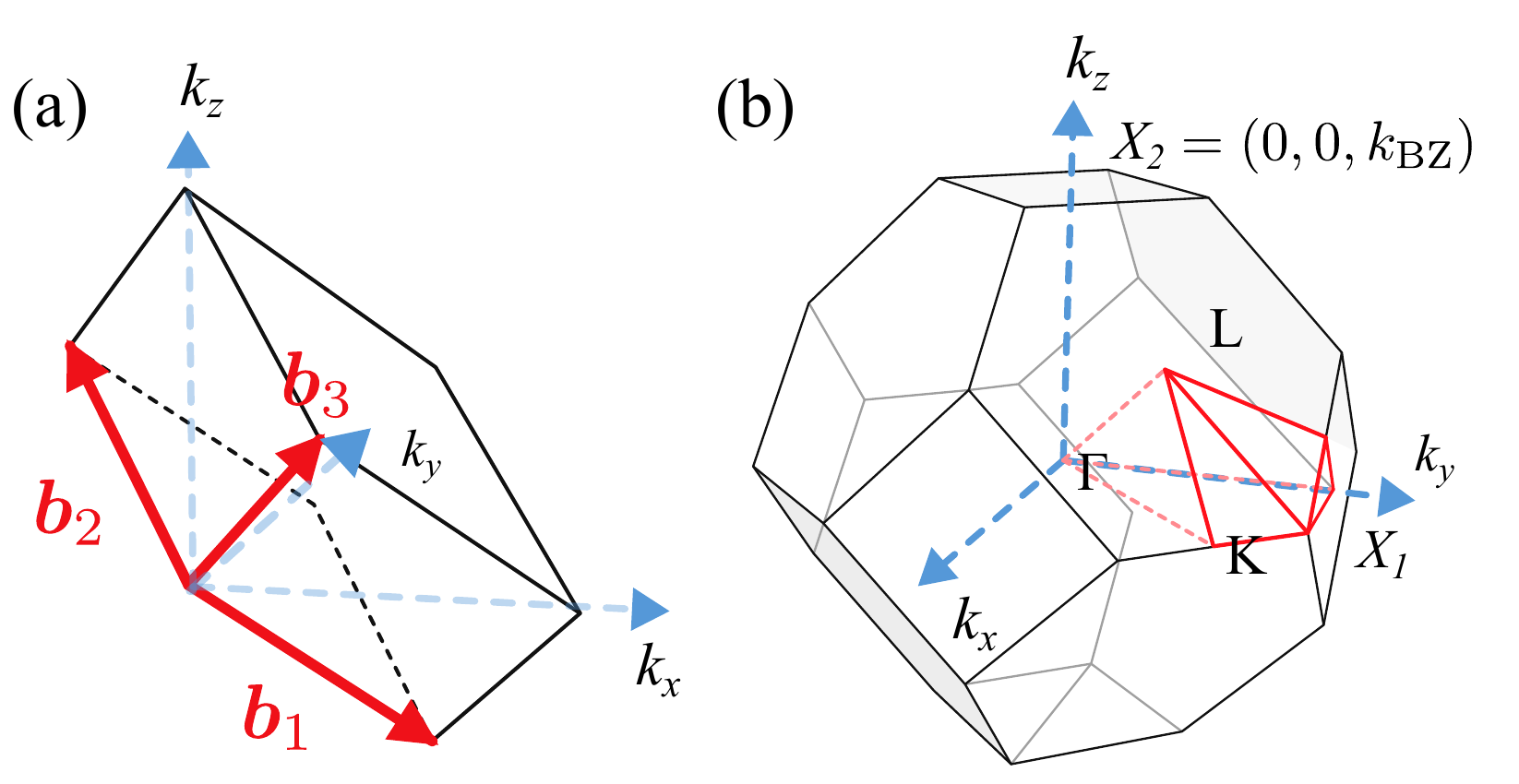}
    \caption{(a) Hexahedron spanned by three reciprocal-lattice vectors ${\bm b}_{1,2,3}$ and (b) the first Brillouin zone (Wigner-Seitz cell) with $\Gamma=(0,\,0\,,\,0)$, ${\rm X}_1 = \pi(0,\,1,\,0)/\sqrt{2}$, ${\rm X}_2 = \pi(0,\,0,\,1)/\sqrt{2}$, ${\rm K} = 3\pi(1,\,1,\,0)/4\sqrt{2}$, and ${\rm L}=\pi(1,\,1,\,1)/2\sqrt{2}$ in the wave vector $\bm k$ space of the pyrochlore magnet. The nearest-neighbor distance $\tilde a$ is set to unity. The value $k_{\rm BZ}=\pi/\sqrt{2}$ represents the boundary of the first Brillouin zone along the $k_z$ direction.}
    \label{fig:cell}
\end{figure}
\section{Magnon and magnon-pair dispersions in a pyrochlore magnet}
\label{sec:dispersion}

In this section, we introduce the pyrochlore lattice and derive the dispersions of its magnon and magnon-pair excitations~\cite{takata2015nematicorderingpyrochloreantiferromagnets}. 
The pyrochlore lattice can be viewed as a three dimensional network of corner-sharing tetrahedra [Fig.~\ref{fig:pyrochlore}(b)].
We use the pyrochlore unit cell shown in Fig.~\ref{fig:pyrochlore}(c). 
The primitive translation vectors of the underlying lattice are $\bm{a}_i=2\bm{e}_i$ ($i=1,2,3$), where $\bm{e}_1=(1,1,0)/\sqrt{2}$, $\bm{e}_2=(0,1,1)/\sqrt{2}$, and $\bm{e}_3=(1,0,1)/\sqrt{2}$ are three nearest-neighbor displacement vectors.
The nearest-neighbor distance $\tilde a$ is set to unity.
The corresponding reciprocal-lattice vectors are $\bm{b}_1 = \pi (1,1,-1)/\sqrt{2}$, $\bm{b}_2 = \pi (-1,1,1)/\sqrt{2}$, and $\bm{b}_3 = \pi (1,-1,1)/\sqrt{2}$, as shown in Fig.~\ref{fig:cell}(a). Representative wave vector points of $\Gamma$, ${\rm X}_1$, ${\rm X}_2$, ${\rm K}$, and ${\rm L}$ are defined in Fig.~\ref{fig:cell}(b). 
An $S=3/2$ antiferromagnetic Heisenberg model supplemented by a biquadratic interaction has been proposed as an effective description of the chromium spinels $A$Cr$_2$O$_4$ ($A=$Cd, Hg, Zn)~\cite{penc2004,ueda2005,ueda2006,miyata2011} on the pyrochlore lattice. 
Its Hamiltonian is~\cite{penc2004}
\begin{equation}
	\mathcal{H}_0 = \sum_{\langle i,j \rangle} \left[J_1 \bm{S}_i \cdot \bm{S}_j + J_2 \left( \bm{S}_i \cdot \bm{S}_j \right)^2 \right] - h \sum_{i} S_i^z \label{eq:CrSpinel}
\end{equation}
where $i$ and $j$ label lattice sites, $\bm{S}_i$ is the spin-$\frac{3}{2}$ operator on site $i$, and the sum $\sum_{\langle i,j \rangle}$ counts each nearest-neighbor bond once. The couplings $J_1>0$ and $J_2<0$ describe nearest-neighbor antiferromagnetic bilinear and ferro-biquadratic exchange, respectively
~\footnote{The biquadratic interaction arises when the displacement-induced modulation of the exchange coupling is combined with the elastic energy that restores the chromium ions to equilibrium. This mechanism yields a negative $J_2$~\cite{Kittel1960}.}, while $h=g\mu_{\mathrm{B}}H$ is the Zeeman energy. Here, $g$, $\mu_{\mathrm{B}}$, and $H$ denote the $g$ factor, the Bohr magneton, and the magnitude of the applied magnetic field. 
The magnetic field is applied along the [001] direction, which we take as the $z$ axis in both real and spin space.
The phase diagram of this model contains several field-induced phases, including a spin-nematic phase generated by magnon-pair condensation~\cite{takata2015nematicorderingpyrochloreantiferromagnets}. Correspondingly, bound magnon pairs coexist with single magnons as low-energy excitations over a broad range of $J_2/J_1$.

We next introduce the DM interaction perturbatively.
Since the DM interaction mixes the magnon and magnon-pair states, magnons and magnon pairs cannot be treated as independent quasiparticles once the DM interaction is fully included.
The DM vectors allowed on the nearest-neighboring bonds on pyrochlore lattice~\cite{Elhajal2005,onose2010} by Moriya's rules~\cite{moriya1960} are
\begin{align}
	\bm{D}_{\mathrm{AB}} &= \frac{D}{\sqrt{2}} (1,-1,0),\, & \bm{D}_{\mathrm{AC}} &= \frac{D}{\sqrt{2}} (0,1,-1), \\  \bm{D}_{\mathrm{AD}} &= \frac{D}{\sqrt{2}} (-1,0,1), \,&
	\bm{D}_{\mathrm{BC}} &= \frac{D}{\sqrt{2}} (1,0,1) , \\   \bm{D}_{\mathrm{BD}} &= \frac{D}{\sqrt{2}} (0,-1,-1),\,& \bm{D}_{\mathrm{CD}} &= \frac{D}{\sqrt{2}} (1,1,0) ,
\end{align}
as illustrated in Fig.~\ref{fig:pyrochlore}(c).
Here, A, B, C, and D denote the indices of the four sublattice sites of the pyrochlore lattice. 
The total Hamiltonian is
\begin{equation}
 \mathcal{H} = \mathcal{H}_0 + \sum_{\langle i,j \rangle}  \bm{D}_{ij} \cdot (\bm{S}_i \times \bm{S}_j) ,
 \label{eq:biquadratic}
\end{equation}
where the DM term supplies the complex hopping amplitudes that endow the magnon and magnon-pair bands with finite Berry curvature.

For magnetic fields above the saturation field, the ground state is given by the fully spin-polarized state $\ket{\mathrm{FP}} \equiv \bigotimes_i \ket{\tfrac{3}{2}}_i$, where $\ket{\tfrac{3}{2}}_i$ is the $S^z=+3/2$ state on site $i$. The ground-state energy is estimated as $E_0 = 27NJ_1/4 + 243NJ_2/16 - 3Nh/2$ in the case without DM interaction, where $N$ is the total number of sites.  
Below, all excitation energies are measured relative to $E_0$.

\subsection{Magnon bands}
\begin{figure}[tbp]
\centering
\includegraphics[width=\linewidth]{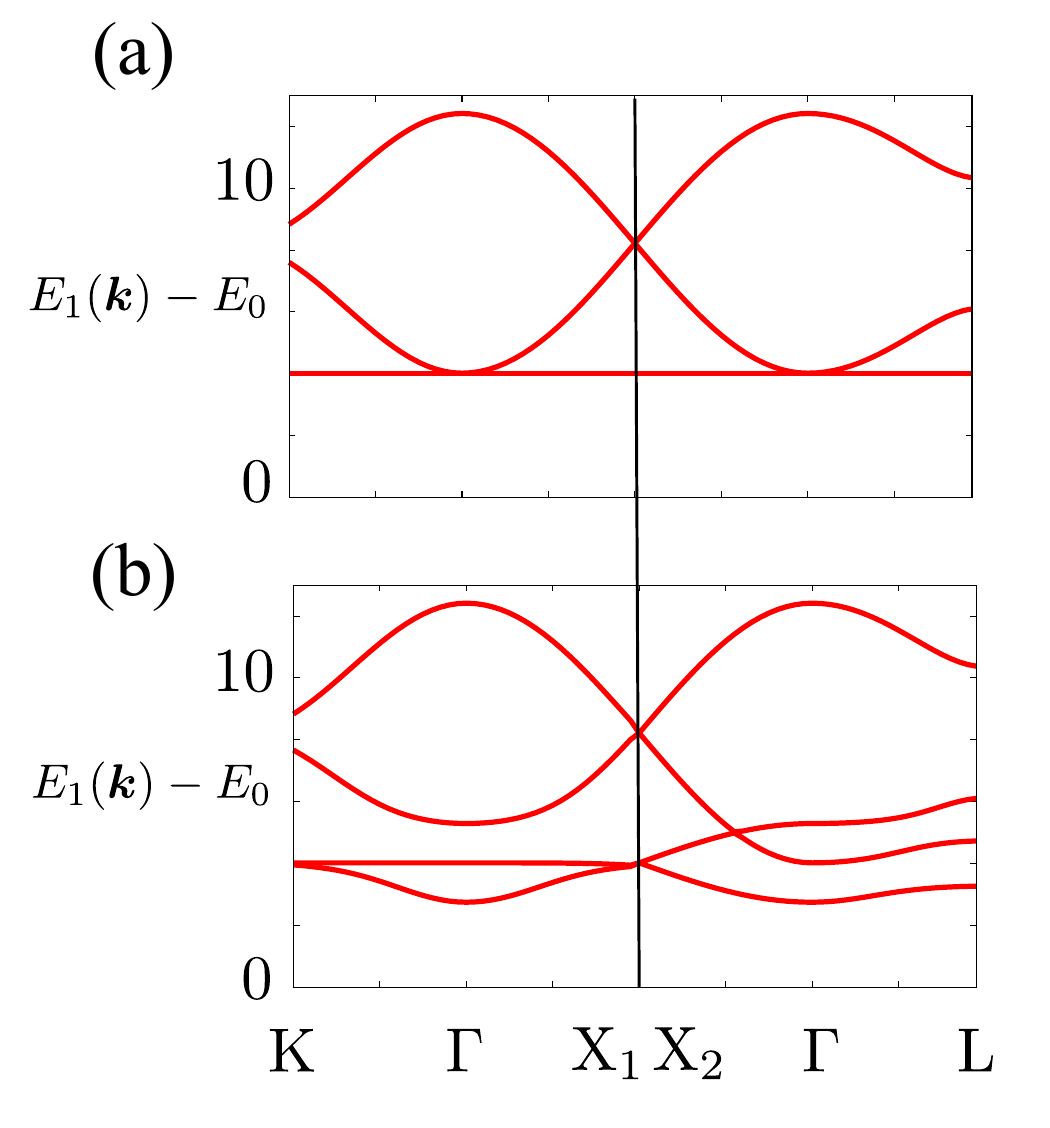}
\caption{Magnon dispersion relations. The vertical axis is the magnon excitation energy $E_1(\bm k) - E_0$ in the saturated state, and the horizontal axis is the wave vector. The left half shows the dispersion along the line segment connecting K-$\mathrm{\Gamma}$-X$_1$, while the right half shows that along X$_2$-$\mathrm{\Gamma}$-L. 
The definition of each wave vector point is shown in Fig.~\ref{fig:cell}(b).
The parameters are $J_2/J_1=-0.2$, $h/J_1=h_{c2}/J_1+1$, with (a) $D/J_1=0$ and (b) $D/J_1=0.3$.}
\label{fig:pyrochlore_magnonSPECTRA}
\end{figure}

First, to compute the energy bands of magnons and magnon pairs, we consider the setup that only the $z$ component of the DM interaction $D_{ij}^z({\bm S}_i\times{\bm S}_j)^z$ is taken into account ($D_{ij}^{x,y}=0$) so that the U(1) spin-rotation symmetry holds and $S^z_{\rm tot}$ is conserved. In this U(1)-symmetric model, a one-magnon state with wave vector $ \bm{k} $ is represented by
\begin{align}
\ket{\alpha(\bm{k})} &\equiv \frac{1}{\sqrt{3N_\mathrm{u}}} \sum_{i \in \alpha} S_i^- e^{i \bm{k} \cdot \bm{r}_i} \ket{\mathrm{FP}} , \label{eq:1mag} 
\end{align}
where $ N_\mathrm{u}=N/4 $ is the number of unit cells and $ \alpha $ (=A, B, C, D) labels the sublattice.
The states $ \ket{\alpha(\bm{k})} $ form an orthonormal basis of the sector with $ S_\mathrm{tot}^z = \frac{3}{2}N-1 $.
The one-magnon energies $ E_1^{(l)}(\bm{k}) $ follow by diagonalizing $ \langle \beta(\bm{k}) | \mathcal{H} | \alpha(\bm{k}) \rangle $.
The Hamiltonian matrix elements are (see App.~\ref{app:pyrochlore_magnon} for details)
\begin{align}
	&\langle \beta(\bm{k})|\mathcal{H}|\alpha(\bm{k})\rangle 
    \nonumber\\&= 
    \begin{cases}
	   \left(3 J_1 + \frac{9}{2} J_2 +3i D_{\alpha \beta}^z \right) \cos (\bm{k} \cdot \bm{\delta}_{\alpha \beta}) & \alpha \neq \beta ,  \\
	    E_0 - 9 J_1 - \frac{27}{2} J_2 + h & \alpha = \beta ,
	\end{cases}
\end{align}
where $\bm{\delta}_{\alpha\beta}$ denotes the nearest-neighbor vector from the $\alpha$ site to the $\beta$ site within the same unit cell, with $\bm{\delta}_{\alpha\beta} = -\bm{\delta}_{\beta\alpha}$.
Figure~\ref{fig:pyrochlore_magnonSPECTRA}(a) shows the dispersion without the DM interaction $D^z_{ij}=0$, whereas Fig.~\ref{fig:pyrochlore_magnonSPECTRA}(b) shows the result for $D/J_1=0.3$.
The remaining parameters are set to $J_2/J_1=-0.2$ and $h/J_1=h_{c2}/J_1+1$, where $h_{c2}$ is the magnetic field at which the magnon-pair band becomes gapless.
With the DM interaction, the band minimum lies at $\bm k = 0$ ($\Gamma$ point), which defines the magnetic critical field of magnons, $h_{c1}$
~\footnote{At $\bm{k}=0$, the energy eigenvalues are $\{h,\, h - 12 J_1 - 18 J_2 ,\, h - 12 J_1 - 18 J_2 - 3 \sqrt{2} D,  \, h - 12 J_1 - 18 J_2 + 3 \sqrt{2} D\}$.}.
Under the realistic condition for weak $J_2$ ($|J_2|/J_1\ll 1$) and $D$ ($|D|/J_1\ll 1$), the critical field is $h_{c1} =  6 (2J_1+3J_2) +3 \sqrt{2} |D|$
~\footnote{Because the DM interaction favors noncollinear spins, it lowers the one-magnon excitation energy and therefore raises the critical field.}: $h_{c1}\simeq 9.67J_1$ for $J_2/J_1=-0.2$ and $D/J_1=0.3$. 
In the absence of the DM interaction, the lowest band is a doubly degenerate flat band, as shown in Fig.~\ref{fig:pyrochlore_magnonSPECTRA}(a).

\subsection{Magnon-pair bands}
The ferro-biquadratic interaction, $J_2<0$, provides an effective attraction that binds two magnons into a pair.
Such a bound pair can propagate through the lattice and can become the lowest excitation near saturation.
For the U(1)-symmetric Hamiltonian with $D_{ij}^{x,y}=0$, a one-magnon state belongs to the sector $S^z_{\mathrm{tot}}=\frac{3}{2}N-1$, whereas a magnon-pair state belongs to $S^z_{\mathrm{tot}}=\frac{3}{2}N-2$. 
The Hamiltonians on these two subspaces can therefore be diagonalized separately unless we introduce the $x$ and $y$ components of the DM interaction.
A general state with two spin flips, including a bound magnon pair ($S^z_{\mathrm{tot}}=S_{\mathrm{tot}}-2$), is a linear combination of $e^{i\bm{k}_1 \cdot \bm{r}_i}  S_i^- e^{i\bm{k}_2 \cdot \bm{r}_j} S_j^- \ket{\mathrm{FP}}$, where the two spin flips have wave vectors $\bm{k}_1$ and $\bm{k}_2$ and are created at sites $i$ and $j$. 
Introducing the center-of-mass momentum $ \bm{k}=\bm{k}_1 + \bm{k}_2 $, relative momentum $ \bm{q}=\left( \bm{k}_1 - \bm{k}_2 \right)/2 $, center-of-mass coordinate $\bm{R}= \left( \bm{r}_i + \bm{r}_j  \right) /2 $, and relative coordinate $\bm{r}=\bm{r}_i - \bm{r}_j  $, we have $\bm{k}_1 \cdot \bm{r}_i + \bm{k}_2 \cdot \bm{r}_j = \bm{k} \cdot \bm{R} + \bm{q} \cdot \bm{r}$.
An exact treatment of the two-magnon sector would retain all relative separations.
However, because both the microscopic interactions and the induced attraction are short-ranged, we approximate the bound state by retaining only on-site and nearest-neighbor pair configurations~\footnote{For each nearest-neighbor configuration with a fixed relative displacement $\bm{\delta}$, the relative momentum $\bm q$ enters only through the phase factor $e^{i\bm q\cdot\bm{\delta}}$ and is therefore not an independent label of the projected real-space basis. Exact two-magnon calculations show that, for sufficiently negative $J_2$, the bound-state weight is concentrated on on-site and nearest-neighbor configurations~\cite{takata2015nematicorderingpyrochloreantiferromagnets}.}.
Under these assumptions, the magnon-pair states are constructed from
\begin{align}
    \nonumber
    &\ket{\alpha \alpha (\bm{k})} \equiv \frac{1}{2 \sqrt{3N_{\mathrm{u}}}} \sum_{i \in \alpha} (S_i^-)^2 e^{i \bm{k} \cdot \bm{r}_i} \ket{\mathrm{FP}},
    \\
    \nonumber
    &\ket{\alpha \beta^{\pm} (\bm{k})} \equiv \frac{1}{3\sqrt{N_\mathrm{u}}} \sum_{i \in \alpha} S_i^- S_{i \pm \bm{\delta}_{\alpha \beta}}^- e^{i \bm{k} \cdot ( \bm{r}_i \pm \frac{\bm{\delta}_{\alpha \beta}}{2} )} \ket{\mathrm{FP}} ,
\end{align}
where the sign $\pm$ distinguishes the two nearest-neighbor $\beta$ sites of a given $\alpha$ site: $\bm{r}_i + \bm{\delta}_{\alpha\beta}$ lies in the same unit cell, whereas $\bm{r}_i - \bm{\delta}_{\alpha\beta}$ lies in an adjacent unit cell.
We refer to the first as an on-site basis state, with two spin flips on one site, and to the second as a bond basis state, with one spin flip on each of two neighboring sites.
Because $\ket{\alpha \beta ^{\pm} (\bm{k})} = \ket{ \beta \alpha ^{\pm} (\bm{k})}$, the bond basis sector contains 12 states: $\ket{\alpha \beta ^{\pm}}= \left\{ \ket{\mathrm{AB} ^{\pm}} , \ket{\mathrm{AC} ^{\pm}}, \ket{\mathrm{AD} ^{\pm}}, \ket{\mathrm{BC} ^{\pm}}, \ket{\mathrm{BD} ^{\pm}}, \ket{\mathrm{CD} ^{\pm}}\right\}$.

\begin{figure}[tbp]
\centering
\includegraphics[width=\linewidth]{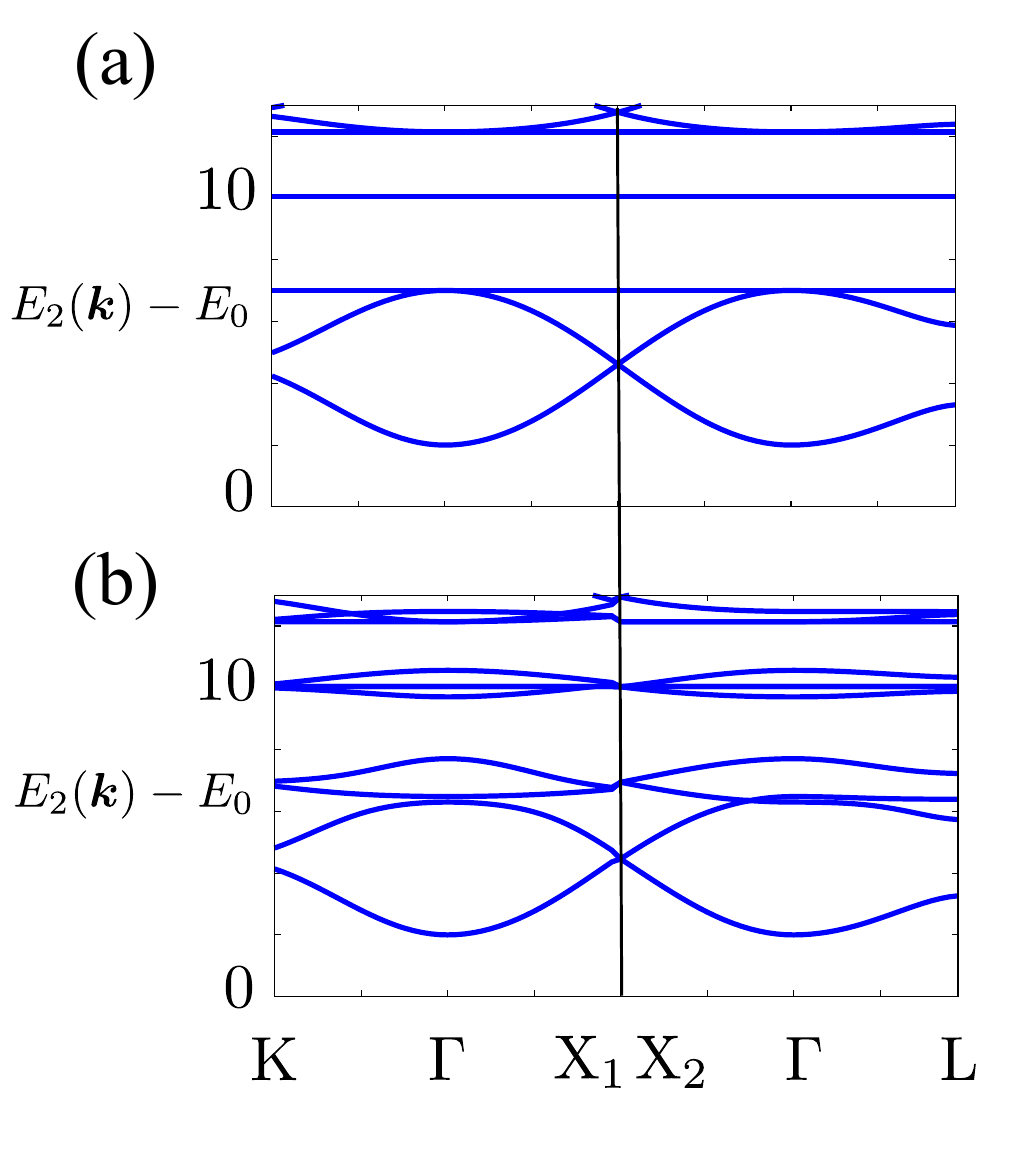}
\caption{Magnon-pair dispersion relations. The vertical axis is the magnon-pair excitation energy $E_2(\bm k)-E_0$ in the saturated state, and the horizontal axis is the wave vector. 
The left half shows the dispersion along the line segment connecting K-$\mathrm{\Gamma}$-X$_1$, while the right half shows that along X$_2$-$\mathrm{\Gamma}$-L. 
The definition of each wave vector point is shown in Fig.~\ref{fig:cell}(b).
The parameters are $J_2/J_1=-0.2$, $h/J_1=h_{c2}/J_1+1$, with (a) $D/J_1=0$ and (b) $D/J_1=0.3$.}
\label{fig:pyrochlore_bi-magnonSPECTRA}
\end{figure}

The projected two-magnon problem is solved by diagonalizing the $16 \times16$ matrix $\langle \gamma \delta (\bm{k})|  \mathcal{H}  |\alpha \beta (\bm{k})\rangle$.
Its nonzero matrix elements are 
\begin{align}
    \label{eq:pair_1}
	&\langle\alpha \alpha (\bm{k})|\mathcal{H}|\alpha \alpha (\bm{k})\rangle =E_0 -18 J_1 - 9 J_2  +2h ,
    \\
	&\langle \beta \beta (\bm{k})|\mathcal{H}|\alpha \alpha (\bm{k})\rangle = 6 J_2 \cos (\bm{k}\cdot \bm{\delta}_{\alpha \beta}) ,
    \\
	&\langle\alpha \beta^{\pm} (\bm{k})|\mathcal{H}|\alpha \alpha (\bm{k}) \rangle 
    \nonumber\\&\hspace{1cm}= \left( \sqrt{3} J_1 -\frac{\sqrt{3}}{2} J_2 + \sqrt{3} i D_{\alpha \beta}^z \right) e^{\mp i \bm{k} \cdot \frac{\bm{\delta}_{\alpha \beta}}{2}  } ,
    \\
	&\langle \alpha \beta^{\pm} (\bm{k})|\mathcal{H}|\alpha \beta^{\pm} (\bm{k}) \rangle =E_0 -17 J_1 - \frac{43}{2} J_2  +2h , \\
	&\langle \alpha \gamma^{\pm} (\bm{k})|\mathcal{H}|\alpha \beta^{\pm} (\bm{k}) \rangle 
    \nonumber\\&\hspace{1cm}= \left( \frac{3}{2} J_1 + \frac{9}{4} J_2 + i \frac{3}{2} D_{\beta\gamma}^z \right) e^{\mp i \bm{k} \cdot \frac{\bm{\delta}_{\beta \gamma}}{2}} ,
    \label{eq:pair_4}
\end{align}
where $\alpha$, $\beta$, and $\gamma$ are mutually distinct (see App.~\ref{app:bi-magnon_matrix} for details).
Diagonalization yields the magnon-pair bands $ E_2(\bm{k}) $.
Panels (a) and (b) of Fig.~\ref{fig:pyrochlore_bi-magnonSPECTRA} show their low-energy dispersions without and with the DM interaction, $D^z_{ij}$, respectively.
In both cases, the lowest magnon-pair band is nondegenerate and has its minimum at $\bm{k}=0$ ($\Gamma$ point).  
Numerical diagonalization gives the pair-condensation field, at which this band becomes gapless, as $h_{c2} \simeq 11.42 J_1$ for $J_2/J_1 = -0.2$, independent of $D$~\footnote{Within the present variational subspace, the minimum magnon-pair energy is independent of the DM interaction. This is likely a consequence of the truncation: Including configurations in which the two magnons are more widely separated may introduce a DM dependence of $h_{c2}$~\cite{takata2015nematicorderingpyrochloreantiferromagnets}.}.

The magnetic field tunes the relative gaps, and hence the thermal populations, of magnons and magnon pairs.
With the field measured from $h_{c2}$ as $x \equiv h/J_1 - h_{c2}/J_1$, the minimum gaps of the magnon and magnon-pair bands, $\Delta_1$ and $\Delta_2$, are given by
\begin{equation}
 \frac{\Delta_1}{J_1} = \frac{h_{c2}-h_{c1}}{J_1} + x, \qquad \frac{\Delta_2}{J_1} = 2x.
 \label{eq:gaps}
\end{equation}
The coefficient 2 in the equation of $\Delta_2$ stems from the fact that a magnon pair has $S^z=-2$ measured from the ground state, while a magnon has $S^z=-1$. For instance, $\Delta_1$ almost coincides with $\Delta_2$ at $h\simeq h_{c2} + 1.8 J_1$ for $J_2/J_1=-0.2$ and $D/J_1=0.3$.

\subsection{Hybridization of magnons and magnon pairs\label{subsect:pyrochlore_mixing}}
\begin{figure}[tbp]
		\centering
		\includegraphics[width=\linewidth]{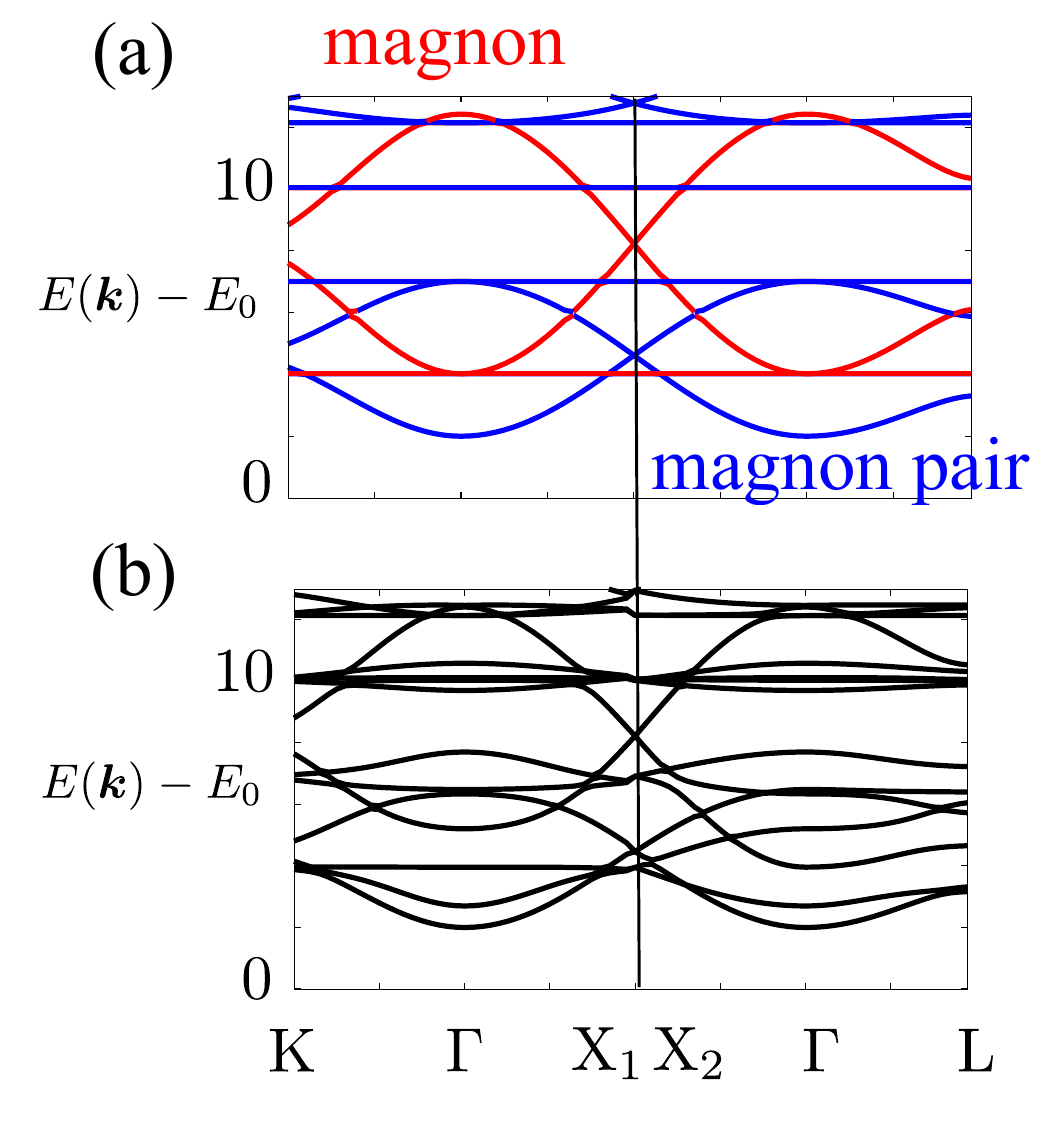}
		\caption{Dispersion relations when magnons and magnon pairs coexist. The excitation energy $E(\bm k)-E_0$ of states in which magnons and magnon pairs are hybridized in the saturated state is shown along lines connecting the high-symmetry points $\mathrm{\Gamma}$, X, K, and L. 
        The definition of each wave vector point is shown in Fig.~\ref{fig:cell}(b).
        The parameters are $J_2/J_1=-0.2$, $h/J_1=h_{c2}/J_1+1$, with (a) $D/J_1=0$ and (b) $D/J_1=0.3$.
        As shown in panel (a), there is no hybridization without $D^{x,y}_{ij}$, and the total band structure simply consists of the magnon and magnon-pair bands. On the other hand, in panel (b), one can no longer distinguish magnons and magnon pairs due to the hybridization.}
\label{fig:pyrochlore_SPECTRA}
\end{figure}

Thus far, we have treated magnons and magnon pairs as independent quasiparticles.
The components of the DM vectors perpendicular to the applied field ($D^{x,y}_{ij}\neq 0$), however, break the U(1) spin-rotation symmetry and couple the one- and two-magnon sectors.
The corresponding off-diagonal block of the Hamiltonian hybridizes the magnon and magnon-pair basis states.
Figure~\ref{fig:pyrochlore_SPECTRA} shows the dispersion obtained by diagonalizing the resulting $20 \times 20$ matrix (see App.~\ref{app:mixing_matrix} for details).
Hybridization opens an avoided-crossing gap where the bare magnon and magnon-pair bands intersect~\footnote{Band hybridization generally concentrates Berry curvature near avoided crossings.}.
Magnons and magnon pairs are no longer well-defined for $D^{x,y}_{ij}\neq 0$ in the mathematical sense, but we can expect that the DM-interaction modified magnon-like (magnon-pair-like) excitations still possess the similar natures to the corresponding magnons (magnon pairs) if the DM interaction is sufficiently small.    

\section{Berry curvature}
\label{sec:Berry}
\begin{figure}[t]
  \centering
  \includegraphics[width=\columnwidth]
    {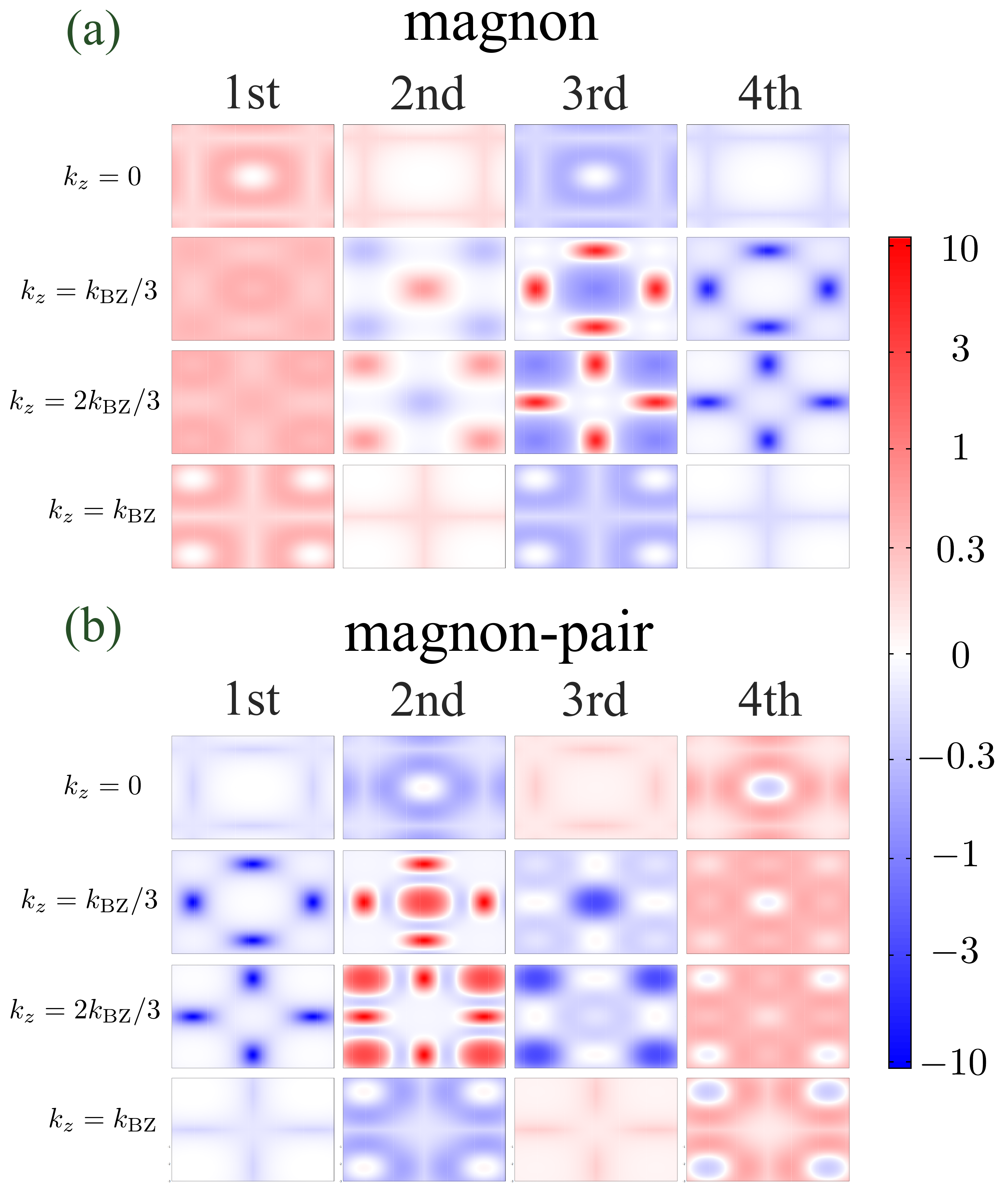}
    \caption{Berry curvature $\Omega^z_n(\mathbf{k})$ on the $k_z=0,\,k_{\mathrm{BZ}}/3,\,2k_{\mathrm{BZ}}/3$, and $k_{\mathrm{BZ}}$ planes for (a) the magnon bands and (b) the magnon-pair bands, where $k_{\mathrm{BZ}}=\pi/\sqrt{2}$ denotes the positive boundary of the first Brillouin zone along the $k_z$ direction. 
    The panels are ordered from lower to higher band energy from left to right. In each panel, the horizontal and vertical axes represent $k_x$ and $k_y$, respectively, and both range from $-3$ to $3$. 
    The magnon and magnon-pair sectors are treated separately, and the parameters are $J_2/J_1=-0.2$ and $D/J_1=0.3$.}
  \label{fig:berry_curvature}
\end{figure}

We next examine the Berry curvatures of the magnon and magnon-pair bands, which govern the intrinsic thermal Hall conductivity. 
The Berry curvature is evaluated numerically using the gauge-invariant discretization introduced by Fukui \textit{et al.}~\cite{Fukui2005}.
Figure~\ref{fig:berry_curvature} shows the Berry curvature on four constant-$k_z$ planes, $k_z=0,\,k_{\mathrm{BZ}}/3,\,2k_{\mathrm{BZ}}/3$, and $k_{\mathrm{BZ}}$, before hybridization between the magnon and magnon-pair sectors is included, where $k_{\mathrm{BZ}}=\pi/\sqrt{2}$ denotes the positive boundary of the first Brillouin zone along the $k_z$ direction, as shown in Fig.~\ref{fig:cell}(b). 
Figure~\ref{fig:berry_curvature} draws the first- to fourth-lowest magnon and magnon-pair bands on each constant-$k_z$ plane. 
The plotted momentum-space region is restricted to $-3\leq k_x,k_y\leq3$. 
We take $J_2/J_1=-0.2$ and $D/J_1=0.3$, as in the dispersion calculations.

A finite DM interaction introduces complex hopping amplitudes into both the magnon and magnon-pair Hamiltonians, thereby endowing their bands with finite Berry curvature.
The lowest magnon and magnon-pair bands carry Berry curvatures of opposite sign throughout the Brillouin zone: positive for the magnon band and negative for the magnon-pair band.
Since these low-energy bands dominate the thermal Hall conductivity at low temperatures, the magnon and magnon-pair contributions are expected to have opposite signs.

As we discussed above, a hybridization between magnons and magnon pairs occurs for $D^{x,y}_{ij}\neq 0$ and one cannot distinguish these two sorts of quasiparticles. However, we numerically verify that the signs of the Berry curvatures of modified magnon and modified magnon-pair bands are unchanged in a broad range of the first Brillouin zone when the DM interaction is small enough.

\section{Thermal Hall conductivity}
\label{sec:thermal_hall}
Matsumoto and Murakami~\cite{Matsumoto2011} formulated the intrinsic thermal Hall response of magnons, including the orbital motion of magnon wave packets.
In this formulation, the thermal transport coefficients are expressed through the Berry curvature $\bm{\Omega}(\bm{k})$ of the magnon Bloch states.
The resulting magnon thermal Hall conductivity, defined through the transverse heat current is
\begin{equation}
	\kappa^{xy} = - \frac{k_{\mathrm{B}}^2 T}{\hbar V} \sum_{n,\bm{k}} c_2\left( \rho_n \right) \Omega_n^z(\bm{k}) \, {\rm [W/Km]}, \label{eq:THC}
\end{equation}
where $n$ labels the magnon bands, $V$ is the system volume, $\Omega_n^{\gamma}(\bm{k})$ is the $\gamma$ component of the Berry curvature on the $n$th band, $k_{\rm{B}}$ is the Boltzmann constant, and $\hbar$ is the reduced Planck constant.
The function $\rho_n \equiv \rho(\varepsilon_{n\bm{k}})$ is the Bose distribution of the $n$th band: $\rho_n=1/\left[ e^{(\varepsilon_{n\bm{k}} - \mu)/k_{\mathrm{B}}T} -1 \right]$, where $\varepsilon_{n\bm{k}}$ is the band energy and $\mu=0$ for thermally excited magnons in equilibrium. 
The weighting function $c_2(\rho)$ is
\begin{align}
\nonumber
	c_2\left( \rho \right) &= \left( 1+\rho \right) \left( \log \frac{1+\rho}{\rho} \right)^2 - \left( \log \rho \right)^2 -2 \mathrm{Li}_2 \left( -\rho \right) ,
\end{align}
where Li$_n (z)$ is the polylogarithm function and can be written as $\mathrm{Li}_2 (z) = -\int_0^{z}  dt \log(1-t)/t $, which we evaluated following Ref.~\cite{Voigt_Polylogarithm_2024}.
Because bosons can macroscopically occupy low-energy states and $c_2$ strongly weights thermally populated modes, the Berry curvature of the lowest bands is especially important.

The same band-geometric formulation applies to other sufficiently long-lived bosonic magnetic quasiparticles.
When magnons and magnon pairs are dilute and their residual interactions are negligible, their energy currents, and therefore their thermal Hall conductivities, are additive. 
Magnon pairs are composite particles: When two pairs spatially overlap, their operators no longer obey canonical bosonic commutation relations. 
In the low-temperature regime considered below, however, the pair density is small, so overlap corrections can be neglected and the pairs can be treated as effective bosons.
We therefore evaluate the total thermal Hall response by using the magnon and magnon-pair band structures, their Berry curvatures and Eq.~(\ref{eq:THC}).

\section{Numerical evaluation of the thermal Hall conductivity}
\label{sec:Numerical}
This section is devoted to the computation of the thermal Hall conductivity in the $S=3/2$ pyrochlore antiferromagnets of Eq.~\eqref{eq:biquadratic} by using Eq.~\eqref{eq:THC}. 
We introduce the dimensionless temperature $t' \equiv k_{\mathrm{B}}T/J_1$ and use the magnetic field detuning $x$ defined in Sec.~\ref{sec:dispersion}.
As shown in Eq.~\eqref{eq:gaps}, the magnetic field tunes the relative thermal populations of magnons and magnon pairs, and hence their competing thermal Hall contributions.
We define the dimensionless conductivity $\tilde{\kappa}^{xy}$ as 
\begin{align}
    \tilde{\kappa}^{xy}\equiv\kappa^{xy}/[k_{\mathrm B}J_1/(\hbar\tilde{a})]
    \label{eq:dimensionless}
\end{align}
For the numerical computation, the Brillouin zone is taken as shown in Fig.~\ref{fig:cell}(a) and it is divided into $101^3$ mesh.
The energy degenerate (band crossing) points of magnon and magnon-pair bands are eliminated numerically. We have checked that the computation results do not quantitatively depend on the mesh size and the small energy window in which the degenerate points are eliminated.

\subsection{U(1)-symmetric case without hybridization}
\begin{figure}[tbp]
\centering
\includegraphics[width=\linewidth]{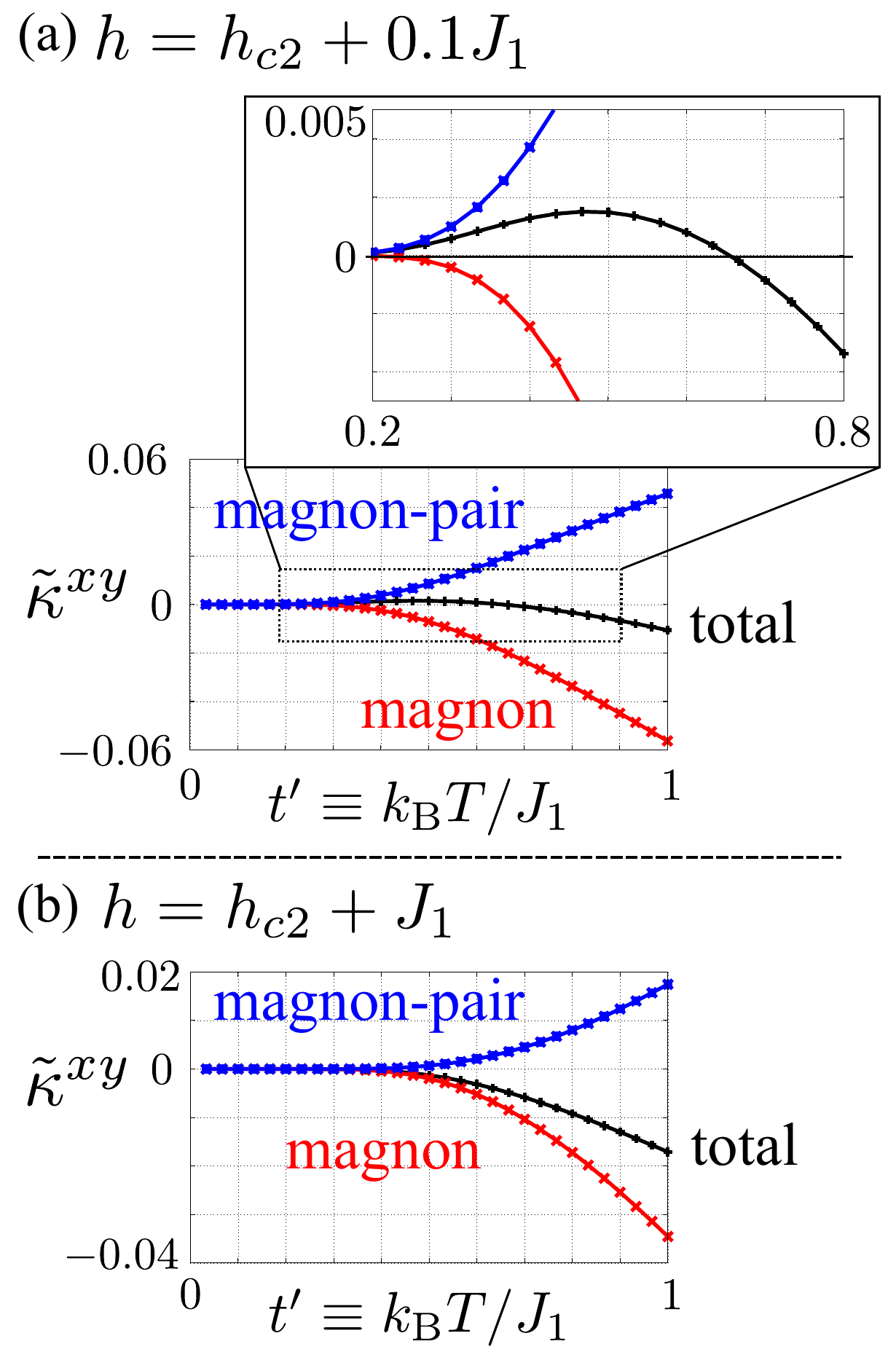}
\caption{Temperature dependence of the magnon and magnon-pair thermal Hall conductivities at magnetic fields (a) $h=h_{c2}+0.1J_1$ ("low" field) and (b) $h_{c2}+J_1$ ("high" field). The red and blue curves represent the magnon and magnon-pair thermal Hall conductivities, respectively, and the black curve is their sum. The sign reversal of the total thermal Hall conductivity is observed in panel (a).}
\label{fig:pyrochlore_wo_mixing_HCvsT}
\end{figure}

First, we focus on the fully polarized phase of the U(1)-symmetric pyrochlore magnets with $D^{x,y}_{ij}=0$, where both magnons and magnon pairs are well defined.  
Figure~\ref{fig:pyrochlore_wo_mixing_HCvsT} shows the temperature dependence of the thermal Hall conductivities, $\tilde\kappa^{xy}$, calculated from the four magnon bands and the four lowest magnon-pair bands.
For the pair sector, we retain only the lowest four bands because we neglected two-magnon states with relative separations beyond nearest neighbors, whose energies are approximately given by the sum of two single-magnon energies, and their mixing with higher-energy magnon-pair bands would undermine the independent magnon-pair quasiparticle picture.
We note that including the fifth and higher projected bands does not appreciably change the calculated conductivity (see App.~\ref{app:other_bands}).

Figure~\ref{fig:pyrochlore_wo_mixing_HCvsT}(a) and (b) show the thermal Hall conductivity as a function of the temperature for a field close to saturation $x=0.1$ and a high field $x=1$, respectively.
The red and blue curves are the magnon and magnon-pair contributions, respectively, and the black curve is their sum.
Over the displayed temperature window, the magnitudes of both contributions grow as the quasiparticle populations increase.
The central result is that the magnon and magnon-pair thermal Hall conductivities have opposite signs; at $x=0.1$, their competition reverses the sign of the total response as the temperature is raised, as shown in the inset. 
This reversal follows directly from the low-energy band geometry and thermal occupation. 
At low temperature, bosonic occupation strongly selects the lowest magnon and magnon-pair bands. 
The relevant component of the Berry curvature is positive throughout the Brillouin zone for the lowest magnon band, whereas it is negative for the lowest magnon-pair band, as shown in Sec.~\ref{sec:Berry}.
The corresponding conductivities are therefore negative and positive, respectively.
For $x=0.1$ ($h=h_{c2}+0.1J_1$), the lowest magnon and magnon-pair bands are well separated in energy~\footnote{The magnon gap is given by $h_{c2}/J_1-h_{c1}/J_1+x$. For $J_2/J_1=-0.2$ and $D/J_1=0.3$, it is $\approx 1.9$. On the other hand, the magnon-pair gap is $2x=0.2$.}, so the lower-lying magnon-pair band dominates the low-temperature response. 
As the temperature increases, the magnon population grows until the total thermal Hall conductivity changes sign.

By contrast, the $x=1$ ($h=h_{c2}+J_1$) result in Fig.~\ref{fig:pyrochlore_wo_mixing_HCvsT}(b) shows no temperature-driven sign reversal. 
At $x=1$, the magnon and magnon-pair gaps are $\min E_1/J_1 \approx 2.8$ and $\min E_2/J_1 = 2$, respectively. 
At this higher magnetic field the lowest magnon and magnon-pair bands are closer in energy, and the larger low-energy Berry curvature of the magnon band prevents the pair contribution from reversing the total response.

\begin{figure}[tbp]
\centering
\includegraphics[width=0.8\linewidth]{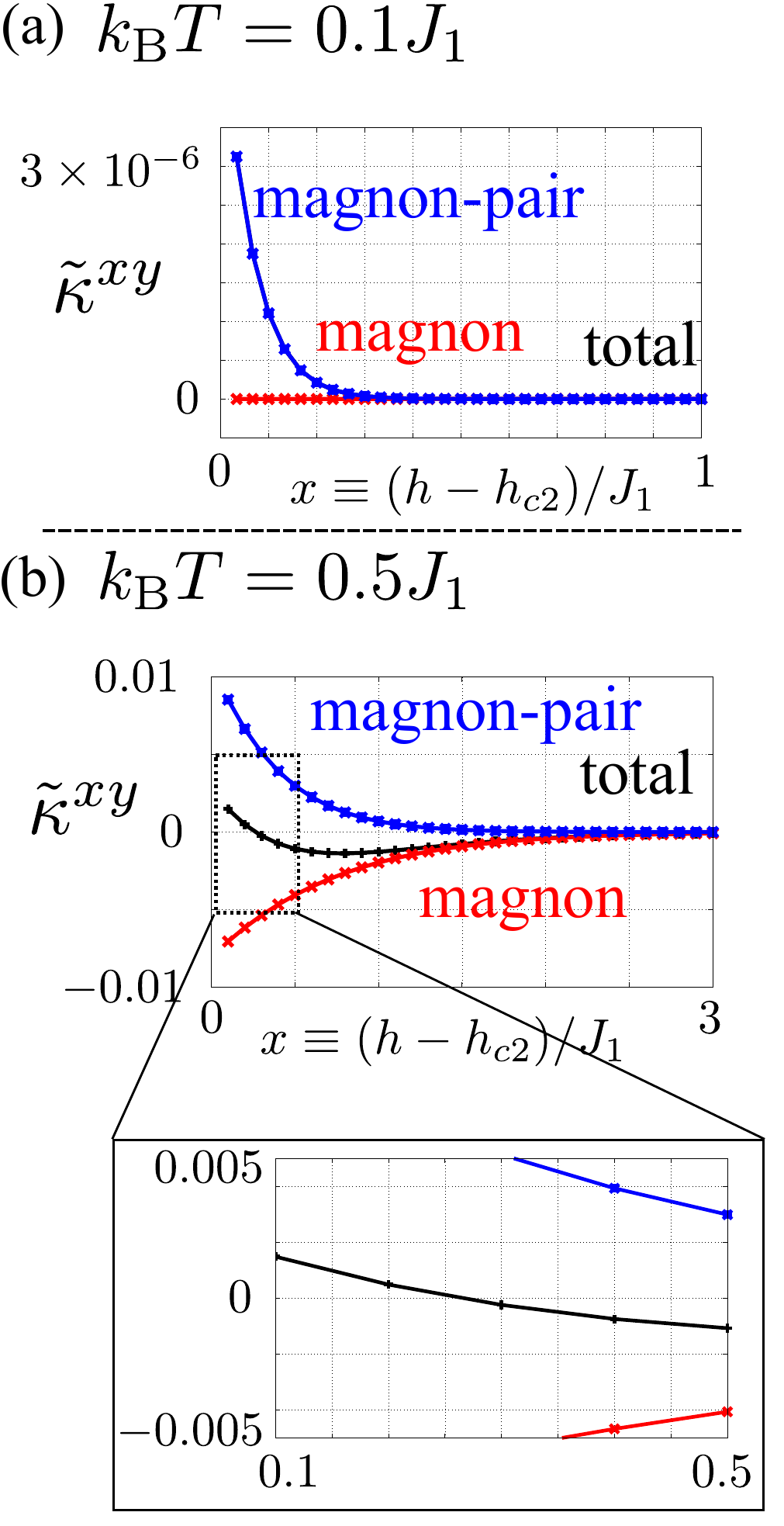}
\caption{Magnetic-field dependence of the magnon and magnon-pair thermal Hall conductivities at temperatures (a) $k_{\rm B}T=0.1J_1$ ($t'=0.1$, low temperature) and (b) $k_{\rm B} T = 0.5 J_1$ ($t'=0.5$, high temperature). The red and blue curves represent the magnon and magnon-pair thermal Hall conductivities, respectively, and the black curve is their sum. The sign reversal of the total thermal Hall conductivity is observed in panel (b).}
\label
{fig:pyrochlore_wo_mixing_HCvsH}
\end{figure}

Figure~\ref{fig:pyrochlore_wo_mixing_HCvsH}(a) and (b) present the thermal Hall conductivity as a function of the magnetic field for the low temperature $t'=0.1$ and the high temperature $t'=0.5$, respectively.
As shown in Fig.~\ref{fig:pyrochlore_wo_mixing_HCvsH}(a), the magnitude of the thermal Hall response grows as $x$ decreases toward the saturation boundary at $t'=0.1$ (low temperature). 
This enhancement reflects the softening of the magnon-pair bands, which increases their thermal occupation.
On the other hand, as shown in Fig.~\ref{fig:pyrochlore_wo_mixing_HCvsH}(b), at $t'=0.5$, both contributions are larger in magnitude than at $t'=0.1$ throughout the plotted field range. 
When the magnetic field is sufficiently close to the saturation field $h_{c2}$, the magnon-pair contribution is dominant like Fig.~\ref{fig:pyrochlore_wo_mixing_HCvsH}(a). As the magnetic field increases, the magnon-pair gap grows twice as rapidly as the single-magnon gap [Eq.~\eqref{eq:gaps}], so the magnon-pair contribution is suppressed more strongly than the magnon contribution. The negative magnon contribution consequently becomes dominant, reversing the sign of the total conductivity, just as in the temperature dependence shown in Fig.~\ref{fig:pyrochlore_wo_mixing_HCvsT}(a).

\subsection{U(1)-broken case with hybridization}
We finally examine the effect of hybridization between the magnon and magnon-pair bands induced by $D^{x,y}_{ij}\neq 0$.
As discussed above, the components of the DM vectors perpendicular to the field break U(1) spin-rotation symmetry and hybridize the two sectors, opening avoided crossings that redistribute and locally enhance the Berry curvature.
If this hybridization were strong, magnons and magnon pairs would no longer be independent quasiparticles, and the decomposition of the thermal Hall conductivity into separate magnon and magnon-pair contributions would lose its meaning.
To examine this, we recalculate the thermal Hall conductivity using the full $20\times20$ projected Hamiltonian.

\begin{figure}[tbp]
	\centering
	\includegraphics[width=0.8\linewidth]{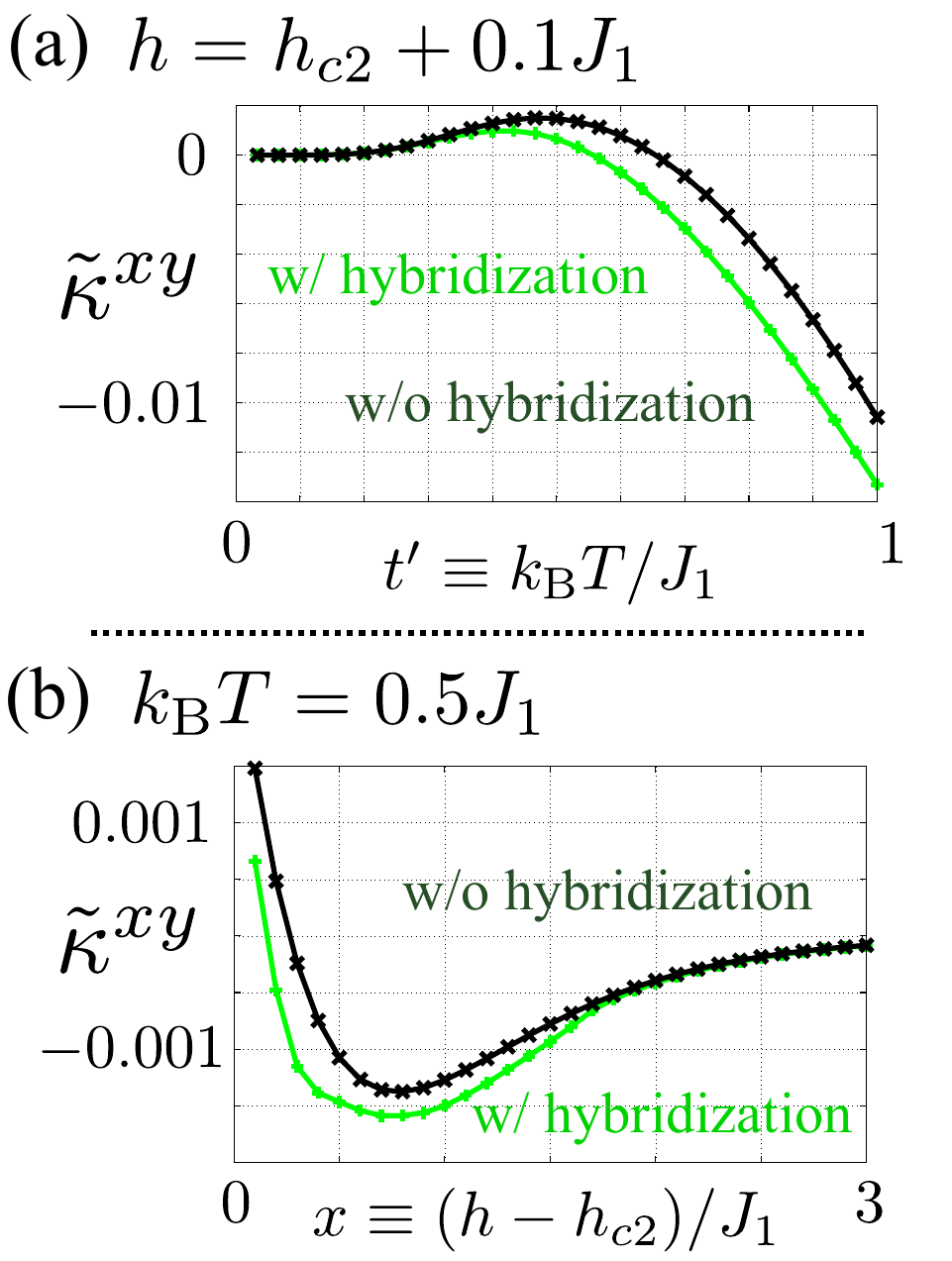}
	\caption{(a) Temperature dependence at $x=0.1$ and (b) magnetic-field dependence at $t'=0.5$ of the total thermal Hall conductivity in the presence and absence of the hybridization. Black and green curves correspond to the thermal Hall conductivity without and with the hybridization, respectively. We set $D/J_1=0.3$ in the case of hybridization.
    }
    \label{fig:pyrochlore_mixing_HCvsT}
\end{figure}

Figure~\ref{fig:pyrochlore_mixing_HCvsT} shows the temperature dependence at $x=0.1$ and the magnetic-field dependence at $t'=0.5$ using the lowest eight bands. 
The strength of the DM interaction is set to $D/J_1=0.3$, same as before. 
The black and green curves show the results without and with hybridization, respectively.
Hybridization preserves both the temperature- and magnetic-field-induced sign reversals.

Let us discuss the deviation between the thermal Hall conductivities with and without the hybridization, shown in Fig.~\ref{fig:pyrochlore_mixing_HCvsT}.
Figure~\ref{fig:pyrochlore_mixing_HCvsT}(a) demonstrates that increasing the temperature enhances the deviation (see around $t'=1.0$).
This deviation enhancement might be understood from the fact that the deformation of magnon and magnon-pair bands due to hybridization is generally larger in the vicinity of level-crossing points compared with the other wave vector regime. 
We note that such level-crossing points are usually far from the lowest excitation energy point of $\rm\Gamma$ [see Fig.~\ref{fig:pyrochlore_SPECTRA}(b)].  
As the temperature increases, the thermal occupation of magnon and magnon-pair states near avoided crossings is increased, thereby the deviation is expected to be enhanced.

For the magnetic-field dependence in Fig.~\ref{fig:pyrochlore_mixing_HCvsT}(b), the deviation persists up to $x\simeq1.5$ and decreases at higher fields. 
This behavior can be interpreted as follows. 
Near the saturation field $h_{c2}$, the lowest magnon and magnon-pair bands are well separated in energy, so their mixing is weak in the U(1)-breaking case. 
Increasing the field reduces their energy separation up to $x\simeq1.8$ [See Eq.~\eqref{eq:gaps} and following statements], where the two gaps coincide, and can therefore enhance the hybridization effect on the band deformation~\footnote{The hybridization involving the lowest magnon-pair state is also suppressed at the ${\rm \Gamma}$ point. At zeroth order in the DM interaction, the matrix elements in Eqs.~\eqref{eq:pair_1}--\eqref{eq:pair_4} make the four on-site configurations symmetry-equivalent, as well as the twelve nearest-neighbor bond configurations. Since the lowest magnon-pair state is nondegenerate, it belongs to the fully symmetric sector and takes the form $|p_{\mathrm{lowest}}(\bm{0})\rangle=A\sum_\alpha|\alpha\alpha(\bm{0})\rangle+B\sum_{\alpha<\beta}\sum_{\sigma=\pm}|\alpha\beta^\sigma(\bm{0})\rangle$. The matrix elements in App.~\ref{app:mixing_matrix} then give $\langle p_{\mathrm{lowest}}(\bm{0})|\mathcal{H}_{\mathrm{DM}}^\perp|\alpha(\bm{0})\rangle=0$, where the identity $\sum_{\beta\ne\alpha}\bm{D}_{\alpha\beta}=0$ is used. This cancellation can further suppress the correction as the crossing between the two lowest bands approaches the ${\rm \Gamma}$ point.}.
When we further increase the magnetic field, both the hybridization effect and the thermal occupation decrease, resulting in the vanishing of the deviation.

We note that (as we mentioned in Sec.~\ref{sec:Berry}) a DM interaction is indispensable for a finite thermal Hall conductivity because Berry curvatures of both the magnons and the magnon pairs vanish without a DM interaction. 
However, an excessively strong DM interaction is generally expected to enhance the hybridization and produce a larger quantitative correction to the fully polarized state and magnetic excitations. 
Such a system with a strong DM interaction should be treated by a nonperturbative method beyond the present perturbative analysis.

\section{Experimental feasibility}
\label{sec:Experiment_Feasibility}
To assess experimental feasibility, we restore physical units and estimate the response for representative chromium spinels.
Since the unit of the thermal Hall conductivity should be ${\rm W/Km}$ in three dimensions, we have to multiply $k_{\rm B} J_1/\hbar\tilde a$ to the numerical results.
For ZnCr$_2$O$_4$, using the lattice constant $a=8.328\,{\rm \AA}$~\cite{Kemei_2013} and antiferromagnetic exchange $J_1=4.5\,{\rm meV}$~\cite{lee2000}, the thermal Hall conductivity, $\kappa^{xy}$, is restored as $\kappa^{xy}\simeq 0.32\tilde\kappa^{xy}\,{\rm W/Km}$ [see Eq.~\eqref{eq:dimensionless}]. 
For HgCr$_2$O$_4$, it becomes $\kappa^{xy}\simeq 0.028\tilde\kappa^{xy}\,{\rm W/Km}$, using $a=8.661$ \AA~\cite{ueda2006} and $J_1=0.41\,{\rm meV}$~\cite{kimura2011}.
We next ask whether the sign-reversed signal near the critical field is large enough to resolve experimentally.
From Fig.~\ref{fig:pyrochlore_wo_mixing_HCvsT}(a) or Fig.~\ref{fig:pyrochlore_wo_mixing_HCvsH}(b), the dimensionless conductivity is approximately $0.002$ at $x=0.1$ and $t'= 0.5$.
For ZnCr$_2$O$_4$, this corresponds to a conductivity of $6.4\times10^{-4}\,{\rm W/Km}$. 
However, when $g=2$, the corresponding field and temperature are approximately $450\,{\rm T}$ and $26\,{\rm K}$ placing the effect in the ultrahigh-field regime above $100\,{\rm T}$.
HgCr$_2$O$_4$ offers a more accessible field scale of approximately $41\,{\rm T}$ at $2.4\,{\rm K}$, and its estimated conductivity is $5.6\times10^{-5}\,{\rm W/Km}$.

For comparison, the experimentally observed magnon thermal Hall conductivity of Lu$_2$V$_2$O$_7$ was approximately $5.0\times 10^{-4}\,{\rm W/Km}$~\cite{onose2010}, and more recently magnon thermal Hall conductivities as small as ($\sim5\times10^{-5}\ {\rm W/Km}$) have been experimentally detected in MnSc$_2$S$_4$~\cite{Takeda2024}.
With regard to the required magnetic field, fields of around 40 T can be generated using steady-field magnets~\cite{Hahn201945} or long-pulse magnets~\cite{Miyata2026}.
Therefore, the predicted signal detection for ZnCr$_2$O$_4$ is experimentally accessible although it requires extremely high magnetic field, while that for HgCr$_2$O$_4$ can be done in realistic magnetic field although it is more difficult to be measured. 
Observing the reversal requires a material that combines a moderate saturation field with a sufficiently large exchange and Berry-curvature-driven response.

\section{Other Candidate systems}
\label{sec:other_candidates}
In this section, we briefly discuss other candidate materials, in which magnons and magnon pairs coexist, in addition to pyrochlore magnets.

Frustrated spin-$1/2$ chain magnets with ferromagnetic nearest-neighbor and antiferromagnetic next-nearest-neighbor exchanges provide a promising setting for bound-magnon transport~\cite{Hirobe2019}. Theoretical studies~\cite{Vekua2007,Kecke2007,Hikihara2008,Sudan2009} have shown that these $J_1$--$J_2$ spin chains can support two-magnon pairing and multi-magnon binding near saturations, with spin-quadrupolar (spin-nematic) or higher-multipolar quasi-long-range order emerging below the saturation field. NMR signatures of magnon pairing (spin-nematic quasi-long-range order) have been proposed theoretically~\cite{Sato2009,Sato2011} and subsequently observed experimentally in several candidate materials~\cite{Nawa2013,Matsui2017,Grafe2017Signatures}.
Furthermore, theoretical studies~\cite{Zhitomirsky_2010,Sato2013} have shown that weak interchain coupling can induce spin-nematic long-range order in higher-dimensional arrays of $J_1$–$J_2$ chains.
Candidate spin-chain compounds include LiCuVO$_4$~\cite{Enderle2005,Mourigal2012,Nawa2013,Hirobe2019}, 
Rb$_2$Cu$_2$Mo$_3$O$_{12}$~\cite{Hase2004,Matsui2017,Hayashida2019}, NaCu$_2$O$_2$~\cite{Drechsler2006,Capogna2005}, LiCu$_2$O$_2$~\cite{Masuda2004,Gippius2004}, and 
LiCuSbO$_4$~\cite{Grafe2017Signatures,Dutton2012}.
For example, a high-field NMR measurement on LiCuVO$_4$ reported NMR peak behavior that is at least qualitatively consistent with spin-nematic order near saturation~\cite{Orlova2017}. 
For the intrinsic thermal Hall mechanism considered here, interchain coupling is required to enable transverse propagation.

In two dimensions, the square-lattice spin-$1/2$ $J_1$--$J_2$ model with competing ferromagnetic and antiferromagnetic exchanges also provides a route to bound-magnon formation and bond-nematic order~\cite{Shannon2006}. 
A square-lattice magnet $\rm BaCdVO(PO_4)_2$~\cite{Nath2008,Skoulatos2019,Bhartiya2019,Ranjith2024} has been extensively investigated as a candidate realization of this frustrated model and a potential host of spin-nematic order.

Another promising example is the Shastry--Sutherland model~\cite{SRIRAMSHASTRY19811069}.
In particular, high-field spectroscopy of SrCu$_2$(BO$_3$)$_2$~\cite{Kageyama1999,Miyahara1999} has provided evidence for the field-induced condensation of bound triplon pairs~\cite{Fogh2024}. 
Triplon bands with DM-interaction-induced Berry curvature have also been studied theoretically and experimentally~\cite{Romhnyi2015Hall,McClarty2017}, and a triplon thermal Hall effect has been predicted~\cite{Romhnyi2015Hall}. 
Its experimental detection, however, has proved challenging, possibly because of interactions among thermally excited triplons~\cite{Suetsugu2022}.

The distorted kagome-lattice magnetic material $\rm Cu_3V_2O_7(OH)_2\cdot 2H_2O$, konwn as the "volborthite", has likewise been extensively investigated as a candidate host of spin-nematic order. Theory predicts that spin-nematic order emerges below the one-third magnetization plateau~\cite{Janson2016}. 
This prediction also implies the coexistence of magnons and magnon pairs in the plateau phase. Experimental studies~\cite{Ishikawa2015,Watanabe2016,Yoshida2017,Kohama2019} have examined a range of observables to elucidate the spin-nematic nature of volborthite.

We emphasize that bound-state formation alone does not guarantee the sign reversal found here. Establishing whether such a reversal occurs requires a material-specific analysis of the excitation spectra and Berry curvatures.

\section{Conclusion}
\label{sec:conclusion}
In summary, we have evaluated the intrinsic thermal Hall conductivity of a saturated pyrochlore antiferromagnet with the DM interaction, in which single magnons and bound magnon pairs coexist as elementary excitations.
Including the pair sector qualitatively changes the total thermal Hall response.
The competition between the oppositely signed magnon and magnon-pair contributions produces sign reversals of the total thermal Hall conductivity upon varying either temperature or magnetic field, as shown in Figs.~\ref{fig:pyrochlore_wo_mixing_HCvsT} and \ref{fig:pyrochlore_wo_mixing_HCvsH}. 
This reversal is a distinctive consequence of multiquasiparticle band geometry (Figs.~\ref{fig:pyrochlore_magnonSPECTRA}--\ref{fig:pyrochlore_SPECTRA}) and Berry curvature (Fig.~\ref{fig:berry_curvature}). 
The DM-induced hybridization between the one- and two-magnon sectors preserves the sign reversals and the overall order of magnitude of the thermal Hall response, while shifting the sign-reversal temperature (Fig.~\ref{fig:pyrochlore_mixing_HCvsT}).
Our estimates in Sec.~\ref{sec:Experiment_Feasibility} show that the signal magnitude can be experimentally relevant in chromium spinels, although simultaneously achieving a strong signal and an accessible saturation field remains challenging.
In Sec.~\ref{sec:other_candidates}, we briefly discuss other candidate materials of the sign reversal. Our proposal of the sign reversal can potentially occur under the condition that magnetic materials possess both magnons and magnon pairs as the low-energy elementary excitations.  
Spin-nematic or nearly spin-nematic ordered magnets often satisfy this condition in their field-induced fully polarized states, and therefore we emphasize that the sign change of thermal Hall conductivity is expected to be observed in a broad class of spin-nematic materials.

A sign reversal with the predicted field and temperature dependence would provide a transport fingerprint of bound magnon pairs coexisting with single magnons.
Our results motivate searches for materials in which bound-state formation, Berry curvature, and moderate saturation fields coincide, thereby making magnon-pair transport both observable and potentially tunable.

Finally, we mention the previous studies for sign reversal of the thermal Hall conductivity by magnetic excitations. 
Besides our proposal, sign reversals in several settings have been predicted or experimentally reported: Thermal Hall conductivities around magnetic phase transitions~\cite{Kasahara2018}, phonon-related mechanisms~\cite{Gillig2023,Nawwar_2025}, or competition among single-magnon bands with opposite Berry curvatures~\cite{Hirschberger2015,Lee2015}.
However, the sign reversal in our setup arises within a single phase (fully polarized phase) from the competition between two distinct species of magnetic quasiparticles, and it can be well tuned by external magnetic field or temperature because of their different Zeeman shifts: A single magnon (magnon pair) changes $S^z_{\mathrm{tot}}$ by one (two).

\begin{acknowledgments}
We thank Tsutomu Momoi, Hosho Katsura and Masaki Oshikawa for valuable discussions about spin nematics and thermal Hall effect. This work is supported by JSPS KAKENHI (Grants No. JP25K07198, No. JP25H02112, No. JP22H05131, No. JP25H01609 and No. JP25H01251) and JST, CREST Grant No. JPMJCR24R5, Japan.
\end{acknowledgments}

\bibliography{ref}

\appendix
\include{Appendix}

\end{document}

%% file: Appendix.tex
\section{Details of the band dispersions}
As shown in Fig.~\ref{fig:pyrochlore_magnonSPECTRA}, the magnon spectrum without a DM interaction contains a flat band of localized-magnon states, a common feature of frustrated lattices~\cite{zhitomirsky2004,zhitomirsky2007}.
Introducing the DM interaction makes the flat band dispersive.
Nevertheless, we show below that both a twofold-degenerate nodal line between the X$_1$ and X$_2$ points and Weyl points along $k_z$ axis survive. In this section, we focus on the case with U(1)-symmetric DM interaction (Namely, there is no hybridization between magnons and magnon pairs). 

First, the nodal line reflects the enhanced symmetry of the pyrochlore Hamiltonian along $\overline{\mathrm{X_1 X_2}}$.
When we consider the $k_z=0$ plane, the X$_1$ points are located at, for example, $(k_x,\,k_y)=(0,\,  \pi/\sqrt{2})$, whereas the X$_2$ points lie at $(k_x,\,k_y)=(\pi/\sqrt{2},\, \pi/\sqrt{2})$.
The Hamiltonian on this $\overline{\mathrm{X_1 X_2}}$ line becomes
\begin{equation}
	\begin{pmatrix}
	A_- & B \\
	B^{\dagger} & A_+
	\end{pmatrix},
\end{equation}
where $\tilde{J} = 2 J_1 + 3 J_2$ and
\begin{align}
	&A_\pm = 
	\begin{pmatrix}
	h-\frac{9}{2} \tilde{J} & \pm \frac{3}{2} \tilde{J} \sin \frac{k_{x}}{\sqrt{2}}  \\
	\pm \frac{3}{2} \tilde{J} \sin \frac{k_{x}}{\sqrt{2}}  & h-\frac{9}{2} \tilde{J}
	\end{pmatrix}, 
    \\
	&B =
	\nonumber\\&\begin{pmatrix}
	0 & \left( -\frac{3iD}{\sqrt{2}} +\frac{3\tilde{J}}{2} \right) \cos \frac{k_{x}}{\sqrt{2}}  \\
	\left( -\frac{3iD}{\sqrt{2}} +\frac{3\tilde{J}}{2} \right) \cos \frac{k_{x}}{\sqrt{2}} & 0
	\end{pmatrix}.
\end{align}
The corresponding energy eigenvalues are
\begin{equation}
	h - \frac{9}{2} \tilde{J} \pm \frac{3}{2} \sqrt{D^2 \left( 1+\cos \left[ \sqrt{2} k_{x} \right] \right)+ \tilde{J}^2 }
\end{equation}
and each branch is certainly doubly degenerate. 
The expression also shows that the energies at the X$_2$ points, $(\pi/\sqrt{2},\, \pi/\sqrt{2})$, are independent of the DM interaction.

Second, we consider the Weyl points. The DM interaction leaves isolated crossings (Weyl points) between the second and third magnon bands at $k_z=\pm\sqrt{2}\arccos[\tilde J/(\tilde J+\sqrt{2}|D|)]$, because (as shown in App.~\ref{app:pyrochlore_magnon}) the energies along the $k_z$ axis are given as
\begin{align}
	&h-3 \tilde{J} \pm 3 \tilde{J} \cos \left[ \frac{k_z}{\sqrt{2}} \right] ,
    \\
    &h - 6 \tilde{J} \pm 3\sqrt{2}D\cos \left[ \frac{k_z}{\sqrt{2}} \right] .
\end{align}

As in the one-magnon spectrum, the magnon-pair bands remain degenerate at the X points both with and without the DM interaction.
The pair spectrum also contains a nodal line along $\overline{\mathrm{X_1 X_2}}$ and Weyl points on the $k_z$ axis.

\section{Contribution of higher-energy bands}
\label{app:other_bands}
In the main text, we evaluate the magnon-pair thermal Hall conductivity using the four lowest bands.
At higher energies, the projected magnon-pair bands may overlap and mix with two-magnon states involving relative separations beyond nearest neighbors. 
Since such mixing makes the magnon-pair quasiparticle picture unreliable, treating the higher-energy projected bands as independent magnon-pair bands is not justified if their contribution to the thermal Hall conductivity is large.

\begin{figure}[bp]
    \centering
    \includegraphics[width=\linewidth]{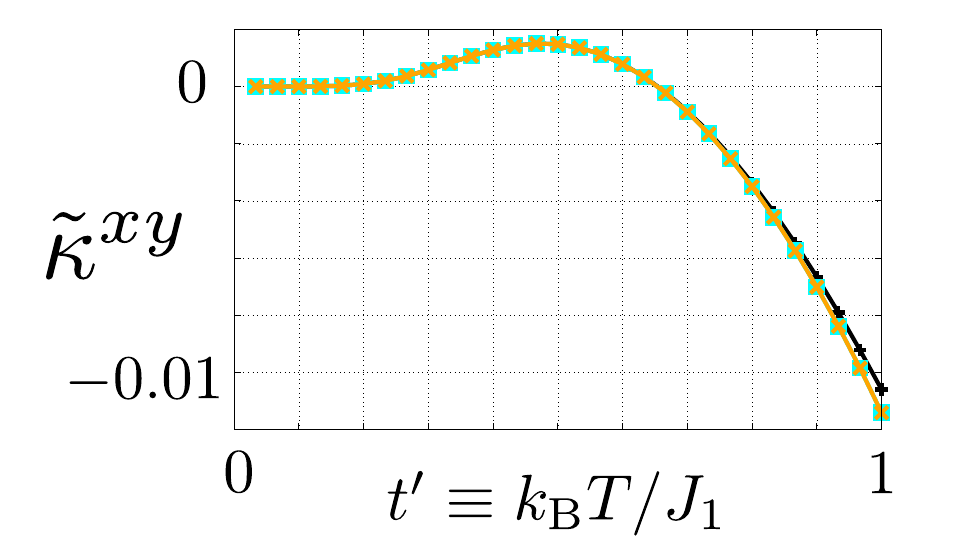}
    \caption{Total thermal Hall conductivity calculated using all four magnon bands and different numbers of magnon-pair bands. The black, orange, and cyan curves include the lowest four, five, and six magnon-pair bands, respectively. The orange and cyan curves are nearly indistinguishable. The parameters are the same as those in Fig.~\ref{fig:pyrochlore_wo_mixing_HCvsT}(a).}
    \label{fig:app:high_energy}
\end{figure}

Figure~\ref{fig:app:high_energy} compares the temperature dependence of the thermal Hall conductivity at $x=0.1$ [same parameters as in Fig.~\ref{fig:pyrochlore_wo_mixing_HCvsT}(a)] obtained with different numbers of magnon-pair bands.
The black curve is the same as the total thermal Hall conductivity in Fig.~\ref{fig:pyrochlore_wo_mixing_HCvsT}(a).
The curves nearly coincide, demonstrating that neglecting the higher-energy projected bands is quantitatively justified.
As discussed in the main text, the Bose occupation and the weighting function $c_2(\rho)$ strongly suppress these higher-energy contributions.

\begin{widetext}
\section{Matrix representation of the magnon Hamiltonian
\label{app:pyrochlore_magnon}}

In this section, we calculate the matrix elements of the magnon Hamiltonian in the U(1)-symmetric case. 
Acting on the one-magnon state $\ket{\alpha(\bm{k})}$, the $J_1$ term of the Hamiltonian gives
\begin{align}
& J_1 \sum_{\langle k,l\rangle} \bm{S}_k \cdot \bm{S}_l \ket{\alpha (\bm{k})} 
\nonumber\\
=& \frac{J_1}{\sqrt{N_{\mathrm u}}} \sum_{i \in \alpha} e^{ i \bm{k} \cdot \bm{r}_i} \sum_{\langle k,l\rangle} \left( \frac{S_k^+ S_l^- + S_k^-  S_l^+}{2} + S_k^z S_l^z \right) \ket{\frac{1}{2}}_i 
\nonumber\\
=& \frac{J_1}{\sqrt{N_{\mathrm u}}} \sum_{i \in \alpha} e^{ i \bm{k} \cdot \bm{r}_i} \sum_{\langle k,l\rangle} \left( 3 \delta_{ik} \ket{\frac{1}{2}}_l  + \left( \frac{3}{2} - \delta_{ik} \right)\left( \frac{3}{2} - \delta_{il} \right) \ket{\frac{1}{2}}_i \right) 
\nonumber\\
=& J_1 \frac{1}{\sqrt{N_{\mathrm u}}} \left\{ \frac{3}{2} \sum_{i \in \alpha} e^{ i \bm{k} \cdot \bm{r}_i}   \sum_{\bm{\delta}_{\alpha}} \ket{\frac{1}{2}}_{i + \bm{\delta_{\alpha}}} + \frac{1}{2} \sum_{i \in \alpha} e^{ i \bm{k} \cdot \bm{r}_i} \left( \frac{27N}{2} - 18 \right) \ket{\frac{1}{2}}_i  \right\} 
\nonumber\\
=& 3 J_1  \sum_{\beta (\neq \alpha)} \cos (\bm{k} \cdot \bm{\delta}_{\alpha\beta}) \ket{\beta(\bm{k})} + J_1  \left( \frac{27}{4}N - 9 \right) \ket{\alpha(\bm{k})} .
\end{align}
$\bm{\delta}_{\alpha}$ denotes a vector from a site on sublattice $\alpha$ to one of its nearest neighbors, while $\bm{\delta}_{\alpha \beta}$ points from the $\alpha$ site to the $\beta$ site within the unit cell and satisfies $\bm{\delta}_{\alpha \beta}=-\bm{\delta}_{ \beta \alpha}$. 
Explicitly, the latter vectors are (Fig.~\ref{fig:pyrochlore})
\begin{align}
\bm{\delta}_{\mathrm{AB}} &= \bm{e}_1 , & \bm{\delta}_{\mathrm{AC}} &= \bm{e}_2 , & \bm{\delta}_{\mathrm{AD}} &= \bm{e}_3 
\nonumber\\
\bm{\delta}_{\mathrm{BC}} &= \bm{e}_2 - \bm{e}_1 , & \bm{\delta}_{\mathrm{BD}} &=  \bm{e}_3 - \bm{e}_1 , & \bm{\delta}_{\mathrm{CD}} &= \bm{e}_3 - \bm{e}_2.
\nonumber
\end{align}
Similarly, the $J_2$ term gives
\begin{align}
& J_2 \sum_{\langle k,l\rangle} \left( \bm{S}_k \cdot \bm{S}_l \right)^2 \ket{\alpha (\bm{k})} 
\nonumber\\
=& J_2\frac{1}{\sqrt{N_{\mathrm u}}} \sum_{i \in \alpha} e^{ i \bm{k} \cdot \bm{r}_i} \sum_{\langle k,l\rangle} \bm{S}_k \cdot \bm{S}_l \left( 3 \delta_{ik} \ket{\frac{1}{2}}_l  + \left( \frac{3}{2} - \delta_{ik} \right)\left( \frac{3}{2} - \delta_{il} \right) \ket{\frac{1}{2}}_i \right) 
\nonumber\\
=&J_2\frac{1}{\sqrt{N_{\mathrm u}}} 3 \sum_{i \in \alpha} e^{ i \bm{k} \cdot \bm{r}_i} \sum_{\langle k,l\rangle}  \delta_{ik} \left( \frac{3}{2} \ket{\frac{1}{2}}_k + \frac{3}{4} \ket{\frac{1}{2}}_l \right)
\nonumber\\
&+ J_2\frac{1}{\sqrt{N_{\mathrm u}}} \sum_{i \in \alpha} e^{ i \bm{k} \cdot \bm{r}_i} \sum_{\langle k,l\rangle}  \left( \frac{9}{4} - \frac{3}{2} \delta_{ik} - \frac{3}{2} \delta_{il} \right)  \left( 3 \delta_{ik} \ket{\frac{1}{2}}_l  + \left( \frac{9}{4} - \frac{3}{2} \delta_{ik} - \frac{3}{2} \delta_{il} \right)  \ket{\frac{1}{2}}_i \right) 
\nonumber\\
=& J_2 \frac{9}{4} \frac{1}{\sqrt{N_{\mathrm{u}}}} \sum_{i \in \alpha} e^{ i \bm{k} \cdot \bm{r}_i} \sum_{l=i+\bm{\delta}_{\alpha}} \ket{\frac{1}{2}}_i + J_2 \frac{9}{8} \frac{1}{\sqrt{N_{\mathrm{u}}}} \sum_{i \in \alpha} e^{ i \bm{k} \cdot \bm{r}_i}   \sum_{l = i + \bm{\delta}_{\alpha}} \ket{\frac{1}{2}}_{l} 
\nonumber\\
&+ J_2\frac{1}{\sqrt{N_{\mathrm u}}} \sum_{i \in \alpha} e^{ i \bm{k} \cdot \bm{r}_i} \sum_{\langle k,l\rangle} \left( \frac{9}{4} \delta_{ik} \ket{\frac{1}{2}}_l + \left( \frac{9}{4} - \frac{3}{2} \delta_{ik}  - \frac{3}{2} \delta_{il} \right)^2 \ket{\frac{1}{2}}_i \right) 
\nonumber\\
=&  J_2 \frac{9}{2} \sum_{\beta (\neq \alpha)} \cos(\bm{k} \cdot \bm{\delta}_{\alpha \beta}) \ket{\beta(\bm{k})}   + J_2   \left( \frac{243}{16}N -  \frac{27}{2} \right) \ket{\alpha(\bm{k})} .
\end{align}
For the DM term, we first rewrite the bond sum as
\begin{align}
&\sum_{\langle k,l\rangle} \bm{D}_{kl} \cdot \left( \bm{S}_k \times \bm{S}_l \right) \ket{\alpha (\bm{k})} 
= \frac{1}{2} \sum_{\beta \neq \gamma} \sum_{k \in \beta} \sum_{l=k \pm \bm{\delta}_{\beta \gamma}} \bm{D}_{\beta \gamma} \cdot \left( \bm{S}_k \times \bm{S}_l \right) \ket{\alpha (\bm{k})} .
\end{align}
Within the one-magnon sector, only the $z$ component of each DM vector contributes. Thus, for a magnetic field along $(0,0,1)$, $\bm{D}_{\mathrm{AB}}$ and $\bm{D}_{\mathrm{CD}}$ do not enter the one-magnon block.
Retaining the $z$ components, we obtain
\begin{align}
&\sum_{\langle k,l\rangle} \bm{D}_{kl} \cdot \left( \bm{S}_k \times \bm{S}_l \right) \ket{\alpha (\bm{k})} 
\nonumber\\
=& \frac{1}{4i} \sum_{\beta \neq \gamma} \sum_{k \in \beta} \sum_{l=k \pm \bm{\delta}_{\beta \gamma}} D_{\beta \gamma}^z   \left( S_k^- S_l^+ - S_k^+ S_l^- \right) \ket{\alpha (\bm{k})} 
\nonumber\\
=& \frac{3}{4i} \frac{1}{\sqrt{N_{\mathrm{u}}}} \sum_{i \in \alpha} e^{i \bm{k} \cdot \bm{r}_i} \sum_{\beta \neq \gamma} \sum_{k \in \beta} \sum_{l=k \pm \bm{\delta}_{\beta \gamma}} D_{\beta \gamma}^z   \left( \delta_{il} \ket{\frac{1}{2}}_k - \delta_{ik} \ket{\frac{1}{2}}_l \right) 
\nonumber\\
=&  \frac{3}{2i} \frac{1}{\sqrt{N_{\mathrm{u}}}} \sum_{i \in \alpha} e^{i \bm{k} \cdot \bm{r}_i} \sum_{\beta (\neq \alpha)} \sum_{k=i \pm \bm{\delta}_{\beta \alpha}} D_{\beta \alpha}^z     \ket{\frac{1}{2}}_k 
= -3i \sum_{\beta(\neq \alpha)} D_{\beta \alpha}^z \cos (\bm{k} \cdot \bm{\delta}_{\alpha \beta}) \ket{\beta(\bm{k})}.
\end{align}
Combining these contributions yields
\begin{align}
&\langle\beta(\bm{k})|\mathcal{H}|\alpha(\bm{k})\rangle 
= \begin{cases}
\left(3 J_1 + \frac{9}{2} J_2 + 3i D_{\alpha\beta}^z \right) \cos (\bm{k} \cdot \bm{\delta}_{\alpha \beta}) & \alpha \neq \beta  ,
\\
E_0 - 9 J_1 - \frac{27}{2} J_2 + h& \alpha = \beta .
\end{cases}
\end{align}

\section{Matrix representation for magnon pairs  \label{app:bi-magnon_matrix}} 
Similarly to the previous section, we here compute the matrix elements of the magnon-pair Hamiltonian in the U(1)-symmetric case. 
We first let the Hamiltonian act on the on-site basis state $\ket{\alpha \alpha (\bm{k})}$. The $J_1$ contribution is
\begin{align}
	&J_1 \sum_{\langle k,l\rangle} \bm{S}_k \cdot \bm{S}_l \ket{\alpha \alpha (\bm{k})} \nonumber \\
	=&J_1 \frac{1}{\sqrt{N_{\mathrm{u}}}} \sum_{i \in \alpha} e^{i \bm{k } \cdot \bm{r}_i} \sum_{\langle k,l\rangle} \left( S_k^+ S_l^- + S_k^z S_l^z \right) \ket{-\frac{1}{2}}_i \nonumber \\
	=& J_1 \frac{1}{\sqrt{N_{\mathrm{u}}}} \sum_{i \in \alpha} e^{i \bm{k } \cdot \bm{r}_i} \sum_{\langle k,l\rangle} \left( 2 \sqrt{3} \delta_{ik} \ket{\frac{1}{2}}_i \ket{\frac{1}{2}}_l + \left( \frac{3}{2} - 2 \delta_{ik} \right) \left( \frac{3}{2} - 2 \delta_{il} \right) \ket{-\frac{1}{2}}_i \right) \nonumber \\
	=& J_1 \frac{1}{\sqrt{N_{\mathrm{u}}}} \sum_{i \in \alpha} e^{i \bm{k } \cdot \bm{r}_i} \left( \sqrt{3} \sum_{\beta (\neq \alpha)} \ket{\frac{1}{2}}_i \ket{\frac{1}{2}}_{i+\bm{\delta}_{\alpha \beta}} + \sqrt{3} \sum_{\beta (\neq \alpha)} \ket{\frac{1}{2}}_i \ket{\frac{1}{2}}_{i - \bm{\delta}_{\alpha \beta}}  + \left( \frac{27}{4}N -18 \right) \ket{-\frac{1}{2}}_i \right) \nonumber \\
	=& J_1  \left( \sqrt{3} \sum_{\beta (\neq \alpha)} e^{-i \bm{k}\cdot \frac{\bm{\delta}_{\alpha \beta}}{2}}  \ket{\alpha \beta ^+ (\bm{k}) } + \sqrt{3} \sum_{\beta (\neq \alpha)} e^{i \bm{k}\cdot \frac{\bm{\delta}_{\alpha \beta}}{2}} \ket{\alpha \beta ^- (\bm{k}) }   + \left( \frac{27}{4}N -18 \right) \ket{\alpha \alpha (\bm{k})} \right).
\end{align}
The $J_2$ contribution is obtained analogously:
\begin{align}
	&J_2 \sum_{\langle k,l\rangle} \left( \bm{S}_k \cdot \bm{S}_l \right)^2 \ket{\alpha \alpha (\bm{k})} \nonumber \\
	=&J_2 \frac{1}{\sqrt{N_{\mathrm{u}}}} \sum_{i \in \alpha} e^{i \bm{k } \cdot \bm{r}_i} \sum_{\langle k,l\rangle} \bm{S}_k \cdot \bm{S}_l  \left( 2 \sqrt{3} \delta_{ik} \ket{\frac{1}{2}}_i \ket{\frac{1}{2}}_l + \left( \frac{3}{2} - 2 \delta_{ik} \right) \left( \frac{3}{2} - 2 \delta_{il} \right) \ket{-\frac{1}{2}}_i \right) \nonumber \\
	=&J_2 \frac{1}{\sqrt{N_{\mathrm{u}}}} 2 \sqrt{3} \sum_{i \in \alpha} e^{i \bm{k } \cdot \bm{r}_i} \sum_{\langle k,l\rangle} \delta_{ik} \left( \sqrt{3} \ket{-\frac{1}{2}}_l + \sqrt{3} \ket{-\frac{1}{2}}_i + \frac{1}{4} \ket{\frac{1}{2}}_i \ket{\frac{1}{2}}_l \right) \nonumber \\
	+&J_2 \frac{1}{\sqrt{N_{\mathrm{u}}}} \sum_{i \in \alpha} e^{i \bm{k } \cdot \bm{r}_i} \sum_{\langle k,l\rangle}    \left( \frac{3}{2} - 2 \delta_{ik} \right) \left( \frac{3}{2} - 2 \delta_{il} \right) \left(  2 \sqrt{3} \delta_{ik} \ket{\frac{1}{2}}_i \ket{\frac{1}{2}}_l + \left( \frac{3}{2} - 2 \delta_{ik} \right) \left( \frac{3}{2} - 2 \delta_{il} \right) \ket{-\frac{1}{2}}_i  \right) \nonumber \\
	=&  6J_2 \sum_{\beta (\neq \alpha)} \cos (\bm{k} \cdot \bm{\delta}_{\alpha \beta})  \ket{\beta \beta (\bm{k})} + 18J_2 \ket{\alpha \alpha (\bm{k})}  + \frac{\sqrt{3}}{4}J_2 \sum_{\beta (\neq \alpha)} \left( e^{-i \bm{k}\cdot \frac{\bm{\delta}_{\alpha \beta}}{2}} \ket{\alpha \beta ^+ (\bm{k})} + e^{i \bm{k}\cdot \frac{\bm{\delta}_{\alpha \beta}}{2}} \ket{\alpha \beta ^- (\bm{k})} \right)  \nonumber \\
	-&\frac{3\sqrt{3}}{4} J_2  \sum_{\beta (\neq \alpha)} \left( e^{-i \bm{k}\cdot \frac{\bm{\delta}_{\alpha \beta}}{2}} \ket{\alpha \beta ^+ (\bm{k})} + e^{i \bm{k}\cdot \frac{\bm{\delta}_{\alpha \beta}}{2}} \ket{\alpha \beta ^- (\bm{k})}  \right) + \left( \frac{243}{16}N - 27\right) J_2 \ket{\alpha \alpha (\bm{k})} \nonumber  \\
	=& \left( \frac{243}{16}N-9\right) J_2 \ket{\alpha \alpha (\bm{k})}+6J_2 \sum_{\beta (\neq \alpha)} \cos (\bm{k} \cdot \bm{\delta}_{\alpha \beta})  \ket{\beta \beta (\bm{k})} \nonumber \\
	-&\frac{\sqrt{3}}{2} J_2  \sum_{\beta (\neq \alpha)} \left( e^{-i \bm{k}\cdot \frac{\bm{\delta}_{\alpha \beta}}{2}} \ket{\alpha \beta ^+ (\bm{k})} + e^{i \bm{k}\cdot\frac{\bm{\delta}_{\alpha \beta}}{2}} \ket{\alpha \beta ^- (\bm{k})}  \right).
\end{align}
For the DM contribution, only the $z$ component of each DM vector contributes:
\begin{align}
	&\sum_{\langle k,l\rangle} \bm{D}_{kl} \cdot \left( \bm{S}_k \times \bm{S}_l \right) \ket{\alpha\alpha (\bm{k})} \nonumber \\
	=& \frac{1}{4i} \sum_{\beta \neq \gamma} \sum_{k \in \beta} \sum_{l=k \pm \bm{\delta}_{\beta \gamma}} D_{\beta \gamma}^z   \left( S_k^- S_l^+ - S_k^+ S_l^- \right) \ket{\alpha \alpha(\bm{k})} \nonumber \\
	=& -\sqrt{3} i \frac{1}{\sqrt{N_{\mathrm{u}}}} \sum_{i \in \alpha} e^{i \bm{k} \cdot \bm{r}_i}  \sum_{\beta (\neq \alpha)} \sum_{k=i \pm \bm{\delta}_{\alpha \beta}} D_{\beta \alpha}^z \ket{\frac{1}{2}}_i \ket{\frac{1}{2}}_k \nonumber  \\
	=& -\sqrt{3} i \sum_{\beta (\neq \alpha)}  D_{\beta \alpha}^z \left( e^{-i \bm{k}\cdot \frac{\bm{\delta}_{\alpha \beta}}{2}} \ket{\alpha \beta ^+ (\bm{k})} + e^{i \bm{k}\cdot \frac{\bm{\delta}_{\alpha \beta}}{2}} \ket{\alpha \beta ^- (\bm{k})} \right).
\end{align}
We next let the Hamiltonian act on a bond basis state.
The $J_1$ contribution is 
\begin{align}
	&J_1 \sum_{\langle k,l\rangle} \bm{S}_k \cdot \bm{S}_l \ket{\alpha \beta ^{\pm} (\bm{k})} 
    = J_1 \frac{1}{\sqrt{N_{\mathrm{u}}}} \sum_{i \in \alpha} e^{i \bm{k} \cdot \left( \bm{r}_i \pm \frac{\bm{\delta}_{\alpha \beta}}{2} \right)} \sum_{\langle k,l\rangle} \left( \frac{S_k^+ S_l^- + S_k^- S_l^+}{2} + S_k^zS_l^z \right) \ket{\frac{1}{2}}_i \ket{\frac{1}{2}}_{i \pm \bm{\delta}_{\alpha \beta}} \nonumber \\
	=&J_1 \frac{1}{\sqrt{N_{\mathrm{u}}}} \sum_{i \in \alpha} e^{i \bm{k} \cdot \left( \bm{r}_i \pm \frac{\bm{\delta}_{\alpha \beta}}{2} \right)}  \sqrt{3} \left( \ket{-\frac{1}{2}}_{i\pm\bm{\delta}_{\alpha \beta}} + \ket{-\frac{1}{2}}_i \right) \nonumber \\
	+& J_1 \frac{1}{\sqrt{N_{\mathrm{u}}}} \sum_{i \in \alpha} e^{i \bm{k} \cdot \left( \bm{r}_i \pm \frac{\bm{\delta}_{\alpha \beta}}{2} \right)} \frac{3}{2} \sum_{\gamma (\neq \alpha, \beta)} \left( \ket{\frac{1}{2}}_{i \pm \bm{\delta}_{\alpha \beta}} \ket{\frac{1}{2}}_{i \pm \bm{\delta}_{\alpha \gamma}} + \ket{\frac{1}{2}}_{i} \ket{\frac{1}{2}}_{i \pm \bm{\delta}_{\alpha \gamma}} \right)\nonumber  \\
	+& J_1 \frac{1}{\sqrt{N_{\mathrm{u}}}} \sum_{i \in \alpha} e^{i \bm{k} \cdot \left( \bm{r}_i \pm \frac{\bm{\delta}_{\alpha \beta}}{2} \right)}  \left( \frac{27}{4} N -17 \right) \ket{\frac{1}{2}}_i \ket{\frac{1}{2}}_{i \pm \bm{\delta}_{\alpha \beta}} \nonumber \\
	=& \sqrt{3} J_1 \left( e^{\mp i \bm{k} \cdot \frac{\bm{\delta}_{\alpha \beta}}{2}} \ket{\beta \beta (\bm{k})} +e^{\pm i \bm{k} \cdot \frac{\bm{\delta}_{\alpha \beta}}{2}} \ket{\alpha \alpha (\bm{k})} \right) \nonumber \\
	+& \frac{3}{2} J_1 \sum_{\gamma (\neq \alpha, \beta)} \left( e^{\mp i \bm{k} \cdot \frac{\bm{\delta}_{\alpha \gamma}}{2} } \ket{\beta \gamma^{\pm} (\bm{k})} + e^{\mp i \bm{k} \cdot \frac{\bm{\delta}_{\beta \gamma}}{2} } \ket{\alpha \gamma^{\pm} (\bm{k})} \right) + \left( \frac{27}{4} N -17 \right) J_1 \ket{\alpha \beta^{\pm} (\bm{k})} ,
\end{align}
where configurations with magnons on next-nearest-neighbor sites have been omitted.
The $J_2$ contribution is 
\begin{align}
	&J_2 \sum_{\langle k,l\rangle} \left( \bm{S}_k \cdot \bm{S}_l \right)^2 \ket{\alpha \beta ^{\pm} (\bm{k})} \nonumber  \\
	=&J_2 \frac{1}{\sqrt{N_{\mathrm{u}}}} \sum_{i \in \alpha} e^{i \bm{k} \cdot \left( \bm{r}_i \pm \frac{\bm{\delta}_{\alpha \beta}}{2} \right)} \sum_{\langle k,l\rangle}  2 \sqrt{3} \left( \delta_{ik} \delta_{i \pm \bm{\delta}_{\alpha \beta},l} + \delta_{i \pm \bm{\delta}_{\alpha \beta},k} \delta_{il} \right) \bm{S}_k \cdot \bm{S}_l \ket{-\frac{1}{2}}_l  \nonumber \\
	+& J_2 \frac{1}{\sqrt{N_{\mathrm{u}}}} \sum_{i \in \alpha} e^{i \bm{k} \cdot \left( \bm{r}_i \pm \frac{\bm{\delta}_{\alpha \beta}}{2} \right)} \sum_{\langle k,l\rangle} \bm{S}_k \cdot \bm{S}_l \left( 3 \delta_{ik}  \left(   1-\delta_{i \pm \bm{\delta}_{\alpha \beta},l} \right) \ket{\frac{1}{2}}_{i \pm \bm{\delta}_{\alpha \beta}} \ket{\frac{1}{2}}_l + 3 \delta_{i \pm \bm{\delta}_{\alpha \beta},k} \left( 1-\delta_{i l} \right) \ket{\frac{1}{2}}_i  \ket{\frac{1}{2}}_l \right)  \nonumber \\
	+&J_2 \frac{1}{\sqrt{N_{\mathrm{u}}}} \sum_{i \in \alpha} e^{i \bm{k} \cdot \left( \bm{r}_i \pm \frac{\bm{\delta}_{\alpha \beta}}{2} \right)} \sum_{\langle k,l\rangle} \left( \frac{3}{2} -\delta_{ik} - \delta_{i \pm \bm{\delta}_{\alpha \beta},k} \right) \left( \frac{3}{2} -\delta_{il} - \delta_{i \pm \bm{\delta}_{\alpha \beta},l} \right) \bm{S}_k \cdot \bm{S}_l \ket{\frac{1}{2}}_i \ket{\frac{1}{2}}_{i \pm \bm{\delta}_{\alpha \beta}}  \nonumber \\
	=&2 \sqrt{3}  J_2 \frac{1}{\sqrt{N_{\mathrm{u}}}} \sum_{i \in \alpha} e^{i \bm{k} \cdot \left( \bm{r}_i \pm \frac{\bm{\delta}_{\alpha \beta}}{2} \right)} \sum_{\langle k,l\rangle}  \left( \delta_{ik} \delta_{i \pm \bm{\delta}_{\alpha \beta},l} + \delta_{i \pm \bm{\delta}_{\alpha \beta},k} \delta_{il} \right) \left( \sqrt{3} \ket{\frac{1}{2}}_k \ket{\frac{1}{2}}_l -\frac{3}{4} \ket{-\frac{1}{2}}_l \right)  \nonumber \\
	+& 3 J_2 \frac{1}{\sqrt{N_{\mathrm{u}}}} \sum_{i \in \alpha} e^{i \bm{k} \cdot \left( \bm{r}_i \pm \frac{\bm{\delta}_{\alpha \beta}}{2} \right)} \sum_{\langle k,l\rangle}  \delta_{ik}  \left(   1-\delta_{i \pm \bm{\delta}_{\alpha \beta},l} \right) \left( \frac{3}{2} \ket{\frac{1}{2}}_{i \pm \bm{\delta}_{\alpha \beta}} \ket{\frac{1}{2}}_k + \frac{3}{4} \ket{\frac{1}{2}}_{i \pm \bm{\delta}_{\alpha \beta}} \ket{\frac{1}{2}}_l \right) \nonumber \\
	+&  3 J_2 \frac{1}{\sqrt{N_{\mathrm{u}}}} \sum_{i \in \alpha} e^{i \bm{k} \cdot \left( \bm{r}_i \pm \frac{\bm{\delta}_{\alpha \beta}}{2} \right)} \sum_{\langle k,l\rangle}  \delta_{i \pm \bm{\delta}_{\alpha \beta},k} \left( 1-\delta_{i l} \right) \left( \frac{3}{2} \ket{\frac{1}{2}}_i  \ket{\frac{1}{2}}_k + \frac{3}{4} \ket{\frac{1}{2}}_i  \ket{\frac{1}{2}}_l \right) \nonumber \\
	+&J_2 \frac{1}{\sqrt{N_{\mathrm{u}}}} \sum_{i \in \alpha} e^{i \bm{k} \cdot \left( \bm{r}_i \pm \frac{\bm{\delta}_{\alpha \beta}}{2} \right)} \sum_{\langle k,l\rangle} \left( \frac{3}{2} -\delta_{ik} - \delta_{i \pm \bm{\delta}_{\alpha \beta},k} \right) \left( \frac{3}{2} -\delta_{il} - \delta_{i \pm \bm{\delta}_{\alpha \beta},l} \right)  \nonumber \\
	&\times \Bigg\{  2 \sqrt{3} \left( \delta_{ik} \delta_{i \pm \bm{\delta}_{\alpha \beta},l} + \delta_{i \pm \bm{\delta}_{\alpha \beta},k} \delta_{il} \right)  \ket{-\frac{1}{2}}_l	+  \left( 3 \delta_{ik}  \left(   1-\delta_{i \pm \bm{\delta}_{\alpha \beta},l} \right) \ket{\frac{1}{2}}_{i \pm \bm{\delta}_{\alpha \beta}} \ket{\frac{1}{2}}_l + 3 \delta_{i \pm \bm{\delta}_{\alpha \beta},k} \left( 1-\delta_{i l} \right) \ket{\frac{1}{2}}_i  \ket{\frac{1}{2}}_l \right)  \nonumber \\
	&+ \left( \frac{3}{2} -\delta_{ik} - \delta_{i \pm \bm{\delta}_{\alpha \beta},k} \right) \left( \frac{3}{2} -\delta_{il} - \delta_{i \pm \bm{\delta}_{\alpha \beta},l} \right)  \ket{\frac{1}{2}}_i \ket{\frac{1}{2}}_{i \pm \bm{\delta}_{\alpha \beta}} \Bigg\}  \nonumber \\
	=& \left( \frac{243}{16} N -\frac{43}{2} \right) J_2 \ket{\alpha \beta ^{\pm } (\bm{k}) } +\frac{9}{4} J_2 \sum_{\gamma (\neq \alpha, \beta)} \left( e^{\mp i \bm{k} \cdot \frac{\bm{\delta}_{\alpha \gamma}}{2}} \ket{\beta \gamma ^{\pm} (\bm{k})} + e^{\mp i \bm{k} \cdot \frac{\bm{\delta}_{\beta \gamma}}{2}} \ket{\alpha \gamma ^{\pm} (\bm{k})} \right) \nonumber \\
	&-\frac{\sqrt{3}}{2} J_2 \left( e^{\pm i \bm{k} \cdot \frac{\bm{\delta}_{\alpha \beta}}{2}} \ket{\alpha \alpha (\bm{k})}  + e^{\mp i \bm{k} \cdot \frac{\bm{\delta}_{\alpha \beta}}{2}} \ket{\beta \beta (\bm{k})}  \right) .
\end{align}
The DM contribution is 
\begin{align}
	&\sum_{\langle k,l\rangle} \bm{D}_{kl} \cdot \left( \bm{S}_k \times \bm{S}_l \right) \ket{\alpha \beta ^{\pm}(\bm{k})}  \nonumber \\
	=& \frac{1}{2i} \frac{1}{\sqrt{N_{\mathrm{u}}}} \sum_{i \in \alpha} e^{i \bm{k} \cdot \left( \bm{r}_i \pm \frac{\bm{\delta}_{\alpha \beta}}{2} \right)} \sum_{\gamma \neq \eta} \sum_{k \in \gamma} \sum_{l = k \pm \bm{\delta}_{\gamma \eta}} D_{ \gamma \eta}^z   \nonumber \\
	&\times \Bigg\{ \sqrt{3} \delta_{il} \left( 2 \delta_{i \pm \bm{\delta}_{\alpha \beta},k} \ket{-\frac{1}{2}}_{i \pm \bm{\delta}_{\alpha \beta}} + \sqrt{3} \left( 1-\delta_{i\pm\bm{\delta}_{\alpha \beta},k} \right) \ket{\frac{1}{2}}_{i \pm \bm{\delta}_{\alpha \beta}} \ket{\frac{1}{2}}_k \right) \nonumber \\
	&\,\,\,\,\,+ \sqrt{3} \delta_{i \pm \bm{\delta}_{\alpha \beta},l} \left( 2 \delta_{i k} \ket{-\frac{1}{2}}_{i} + \sqrt{3} \left( 1-\delta_{ik} \right) \ket{\frac{1}{2}}_{i } \ket{\frac{1}{2}}_k \right) \Bigg\}  \nonumber \\
	=& \frac{1}{2i} \frac{1}{\sqrt{N_{\mathrm{u}}}} \sum_{i \in \alpha} e^{i \bm{k} \cdot \left( \bm{r}_i \pm \frac{\bm{\delta}_{\alpha \beta}}{2} \right)}   \nonumber \\
	&\times \Bigg\{   2\sqrt{3} D_{\beta \alpha}^z \ket{-\frac{1}{2}}_{i \pm \bm{\delta}_{\alpha \beta}} + 3 \sum_{\gamma(\neq \alpha , \beta)} D_{\gamma \alpha}^z \ket{\frac{1}{2}}_{i \pm \bm{\delta}_{\alpha \beta}} \ket{\frac{1}{2}}_{i\pm \bm{\delta}_{\alpha \gamma}} + 2\sqrt{3} D_{ \alpha \beta}^z \ket{-\frac{1}{2}}_{i} + 3 \sum_{\gamma(\neq \alpha , \beta)} D_{\gamma \beta}^z \ket{\frac{1}{2}}_{i} \ket{\frac{1}{2}}_{i\pm \bm{\delta}_{\alpha \gamma}} \Bigg\}  \nonumber \\
	=& -i \sqrt{3} D_{\alpha \beta}^z \left( e^{\mp i \bm{k} \cdot \frac{\bm{\delta}_{ \beta \alpha}}{2} } \ket{\alpha \alpha (\bm{k})} - e^{\mp i \bm{k} \cdot \frac{\bm{\delta}_{\alpha \beta}}{2} } \ket{\beta \beta (\bm{k})} \right) - i \frac{3}{2} \sum_{\gamma (\neq \alpha, \beta)} \left( D_{\gamma \alpha}^z e^{\mp i \bm{k} \cdot \frac{\bm{\delta}_{\alpha \gamma}}{2}} \ket{\beta \gamma^{\pm} (\bm{k})} +D_{\gamma \beta}^z e^{\mp i \bm{k} \cdot \frac{\bm{\delta}_{\beta \gamma}}{2}} \ket{\alpha \gamma^{\pm} (\bm{k})} \right).
\end{align}
Collecting all contributions, we obtain
\begin{align}
	\langle\alpha \alpha (\bm{k})|\mathcal{H}|\alpha \alpha (\bm{k})\rangle &= J_1 \left(\frac{27}{4} N -18 \right) + J_2 \left( \frac{243}{16}N -9 \right) -h\left( \frac{3}{2} N -2 \right) ,\\
	\langle\beta \beta (\bm{k})|\mathcal{H}|\alpha \alpha (\bm{k})\rangle &= 6 J_2 \cos (\bm{k}\cdot \bm{\delta_{\alpha \beta}}) ,\\
	\langle\alpha \beta^{\pm} (\bm{k})|\mathcal{H}|\alpha \alpha (\bm{k}) \rangle &= \left( \sqrt{3} J_1 -\frac{\sqrt{3}}{2} J_2 + \sqrt{3} i D_{\alpha \beta}^z \right) e^{\mp i \bm{k} \cdot \frac{\bm{\delta}_{\alpha \beta}}{2}  },\\
	\langle \alpha \beta^{\pm} (\bm{k})|\mathcal{H}|\alpha \beta^{\pm} (\bm{k}) \rangle &= J_1 \left(\frac{27}{4} N -17 \right) + J_2 \left( \frac{243}{16}N - \frac{43}{2} \right) -h\left( \frac{3}{2} N -2 \right) ,\\
	\langle \alpha \gamma^{\pm} (\bm{k})|\mathcal{H}|\alpha \beta^{\pm} (\bm{k}) \rangle &= \left( \frac{3}{2} J_1 + \frac{9}{4} J_2 + i \frac{3}{2} D_{\beta\gamma}^z \right) e^{\mp i \bm{k} \cdot \frac{\bm{\delta}_{\beta \gamma}}{2}}.
\end{align}
where $\alpha$, $\beta$, and $\gamma$ are mutually distinct; all other matrix elements vanish.

\section{Hybridization Terms Induced by the DM Interaction\label{app:mixing_matrix}}

In this final Appendix, we consider the DM interaction induced hybridization terms between a magnon and a magnon-pair states in the U(1)-breaking model. 
Acting with the $x$ component of the DM interaction on the magnon state $\ket{\alpha(\bm{k})}$ gives
\begin{align}
 \sum_{\langle k,l\rangle} D_{kl}^x \left( S_k^y S_l^z - S_k^z S_l^y \right) \ket{\alpha (\bm{k})} 
=  \frac{1}{4i} \sum_{\beta \neq \gamma} \sum_{k \in \beta} \sum_{l=k\pm \bm{\delta}_{\beta \gamma}} D_{\beta \gamma}^x \left( S_k^+ S_l^z - S_k^- S_l^z  - S_k^z S_l^+  + S_k^z S_l^-  \right) \ket{\alpha (\bm{k})}.
\end{align}
Using the pyrochlore identity $\sum_{\beta(\neq \alpha)} \bm{D}_{\alpha \beta} = 0$, the surviving terms are
\begin{align}
\frac{1}{4i} \sum_{\beta \neq \gamma} \sum_{k \in \beta} \sum_{l=k\pm \bm{\delta}_{\beta \gamma}} D_{\beta \gamma}^x \left( - S_k^- S_l^z   + S_k^z S_l^-  \right) \ket{\alpha (\bm{k})} 
= - \frac{\sqrt{3}}{2i} \sum_{\beta (\neq \alpha)} D_{\alpha\beta}^x \left( e^{- i\bm{k} \cdot \frac{\bm{\delta}_{\alpha \beta}}{2}} \ket{\alpha \beta ^+ (\bm{k})} +  e^{ i\bm{k} \cdot \frac{\bm{\delta}_{\alpha \beta}}{2}} \ket{\alpha \beta ^- (\bm{k})} \right).
\end{align}
The $y$ component of the DM interaction acts similarly:
\begin{align}
& \sum_{\langle k,l\rangle} D_{kl}^y \left( S_k^z S_l^x - S_k^x S_l^z \right) \ket{\alpha (\bm{k})} 
= - \frac{\sqrt{3}}{2} \sum_{\beta (\neq \alpha)} D_{\alpha\beta}^y \left( e^{- i\bm{k} \cdot \frac{\bm{\delta}_{\alpha \beta}}{2}} \ket{\alpha \beta ^+ (\bm{k})} +  e^{ i\bm{k} \cdot \frac{\bm{\delta}_{\alpha \beta}}{2}} \ket{\alpha \beta ^- (\bm{k})} \right).
\end{align}
Combining the two components yields
\begin{equation}
\langle\alpha \beta ^{\pm} (\bm{k})|\mathcal{H}|\alpha(\bm{k})\rangle =  \frac{\sqrt{3}}{2} \left( i D_{ \alpha \beta} ^x  - D_{ \alpha \beta}^y \right)e^{\mp i \bm{k} \cdot \frac{\bm{\delta}_{\alpha \beta}}{2}}.
\end{equation}
\end{widetext}